\documentclass[12pt]{elsart}

\usepackage{graphicx}
\usepackage{subfigure}
\usepackage{amsmath}
\usepackage{cite}
\usepackage{tikz}
\usepackage{setspace}

\DeclareMathOperator*{\argmin}{arg\,min}

\journal{Physics Letters A}

\begin{document}

\begin{frontmatter}

\title{Entropy of spin clusters with frustrated geometry}
\author{M. \v{Z}ukovi\v{c}\corauthref{cor}},
\ead{milan.zukovic@upjs.sk}
\author{A. Bob\'ak}
\ead{andrej.bobak@upjs.sk}
\address{Department of Theoretical Physics and Astrophysics, Faculty of Science,\\ 
P. J. \v{S}af\'arik University, Park Angelinum 9, 041 54 Ko\v{s}ice, Slovakia}
\corauth[cor]{Corresponding author.}

\begin{abstract}
Geometrically frustrated clusters of Ising spins of different shapes on a triangular lattice are studied by exact enumeration and Monte Carlo simulation. The focus is laid on the ground-state energy and residual entropy behaviors as functions of the cluster shape and size, as well as the spin value. Depending on the cluster shape, the residual entropy density on approach to the thermodynamic limit can either vanish or remain finite and the dependence can be decreasing, increasing or non-monotonic. Nevertheless, the relative entropies normalized by the respective thermodynamic limit values turn out to be little sensitive to the spin value. Some interesting finite-temperature properties of the clusters are also discussed and attention is drawn to their magnetocaloric properties.   
\end{abstract}


\end{frontmatter}

\section{Introduction}
\hspace*{5mm} 
The effect of geometrical frustration in spin systems is a subject of intensive investigations. One of the simplest such models is a triangular lattice Ising antiferromagnet (TLIA) with spin $s=1/2$, which shows no long-range ordering at any finite temperature~\cite{wann,step}. TLIA is fully frustrated and the ground state (GS) is highly degenerated with non-vanishing entropy~\cite{wann}. The lack of order is due to large ground-state degeneracy, nevertheless, the latter can be lifted by various perturbations, such as an external magnetic field~\cite{metcalf,schick,netz,zuko} or selective dilution~\cite{kaya,zuko1}. 
The degeneracy can also be removed in case of finite lattices and certain types of boundary conditions~\cite{griff,aizen}. By increasing the lattice size to infinity the ground-state entropy in the thermodynamic limit is recovered only if the boundary conditions giving maximum degeneracy are considered~\cite{aizen,simon}. For example, if the lattice has a shape of a rhombus the ground state is non-degenerate (unique up to a global spin inversion) for any lattice size and thus the entropy density vanishes in the thermodynamic limit~\cite{milla1}. Millane et al.~\cite{milla2,milla3} have investigated the dependence of the entropy density on different boundary conditions on finite triangular domains for which the ground states do not admit a dimer covering and found that it can vary between zero and maximal value.\\
\hspace*{5mm} Several studies~\cite{naga,yama,lipo,zeng,zuko2} have shown that the ground-state properties of TLIA can be significantly affected by the spin value $s$. For sufficiently large $s$ even a long-range order can occur as a result of the presence of so-called free spins (their state does not affect the energy), the degeneracy of which then outweighs the degeneracy of the disordered phase~\cite{lipo}. Naturally, one can expect that the spin value can also affect the ground-state energies and degeneracies of finite clusters of different shapes. To our best knowledge the character of such dependence has not been studied yet. \\
\hspace*{5mm} In the present paper we consider geometrically frustrated Ising spin-$s$ ($s=1/2,1$ and $3/2$) clusters of various shapes and sizes with free boundaries and study effects of the shape, size and the spin value on the ground-state energy and residual entropy. In particular, we are interested in how these quantities can be influenced by minimal manipulation in the clusters' shapes, such as removal of a small number of vertices. We study their behavior for different $s$ on approach to the respective thermodynamic limits, which for $s>1/2$ are obtained by Monte Carlo simulations~\cite{zuko2}. Some interesting finite-temperature properties of the clusters with $s=1/2$ are also discussed.

\section{Model and methods} 
\begin{figure}[t]
\centering
    \subfigure[R]{\includegraphics[scale=0.3,clip]{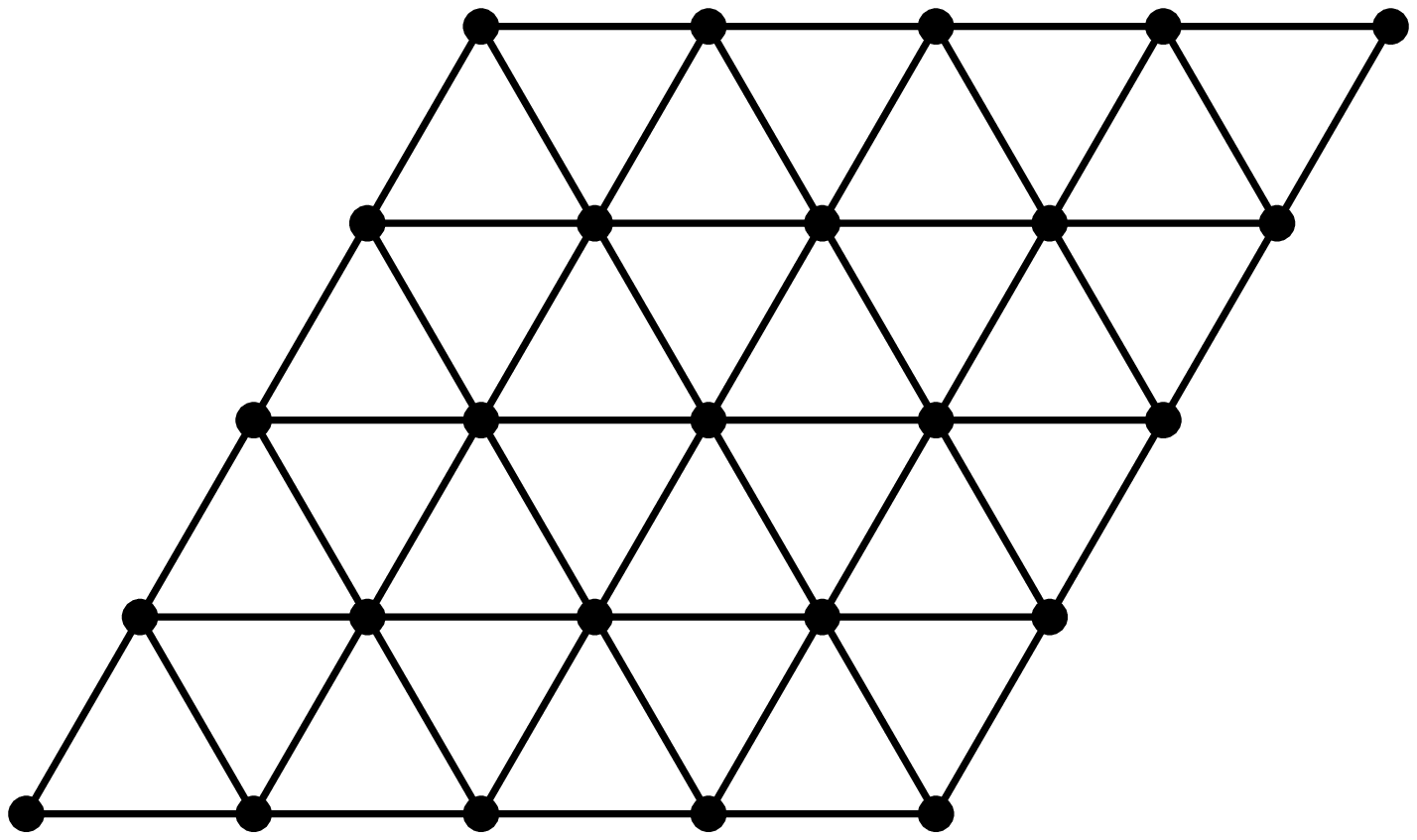}\label{fig:rhomb}}
    \subfigure[R1]{\includegraphics[scale=0.3,clip]{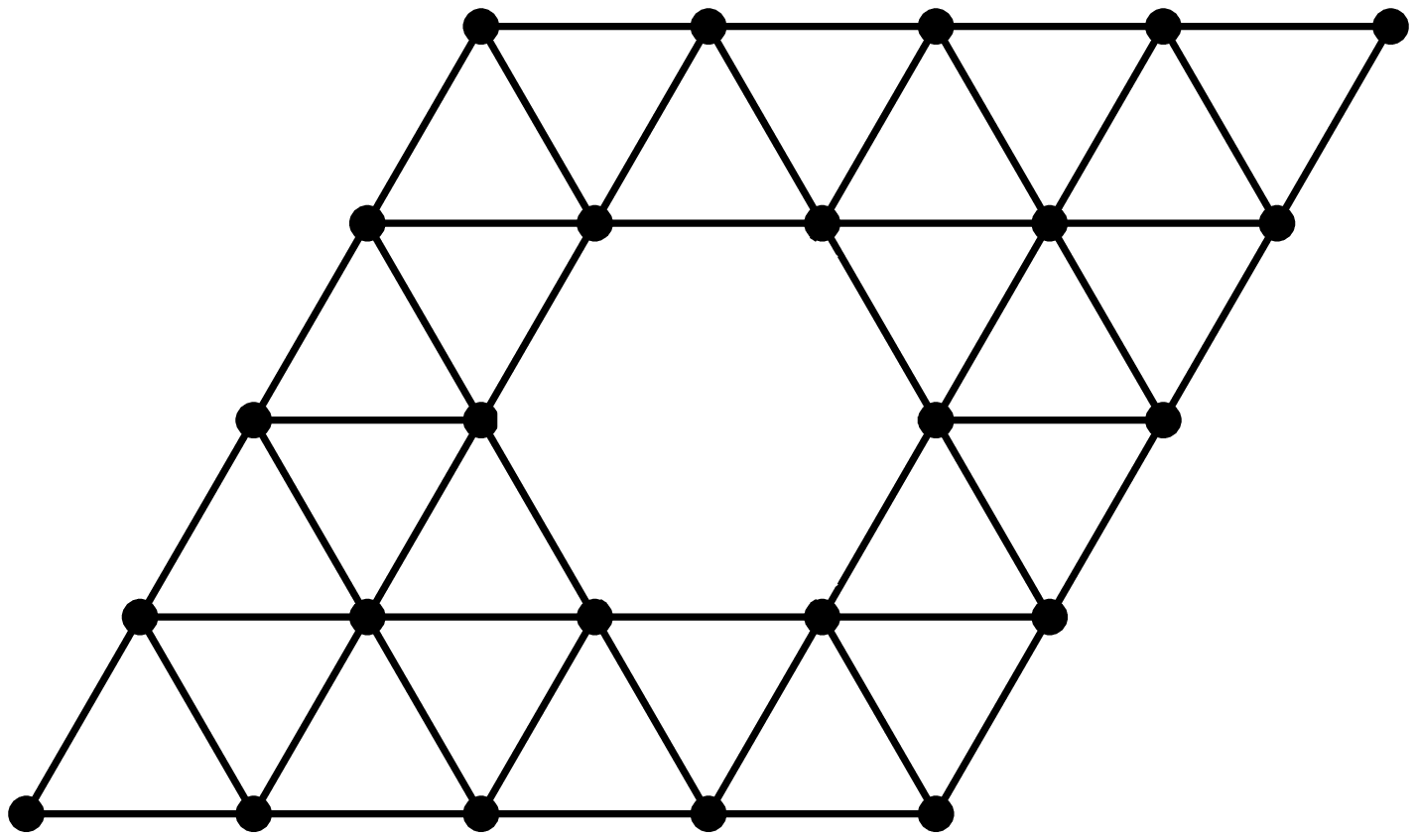}\label{fig:rhomb1}} \\
    \subfigure[H]{\includegraphics[scale=0.3,clip]{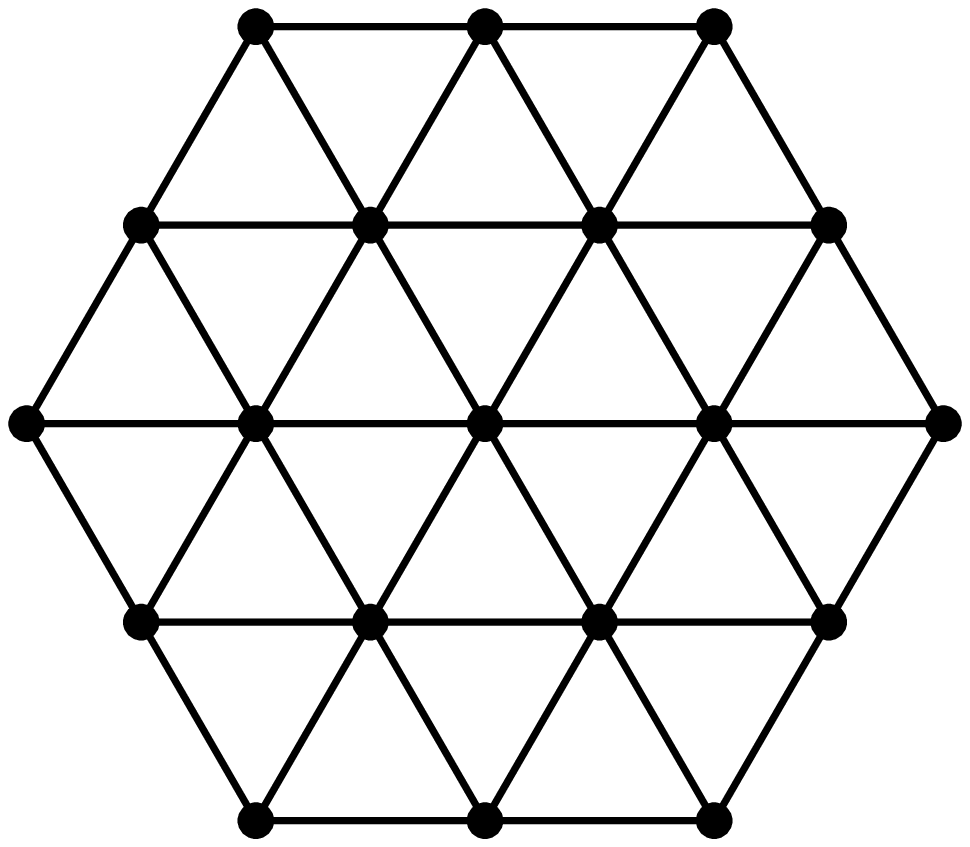}\label{fig:hexa}} 
    \subfigure[H1]{\includegraphics[scale=0.3,clip]{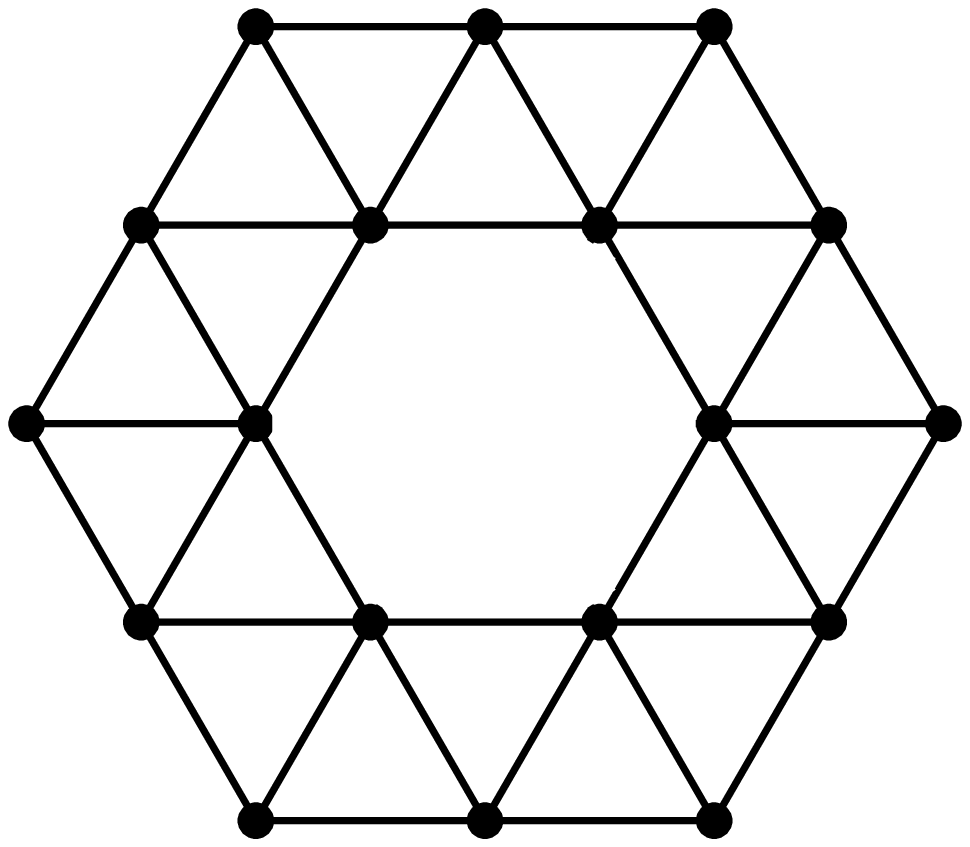}\label{fig:hexa1}} \\
    \subfigure[T]{\includegraphics[scale=0.3,clip]{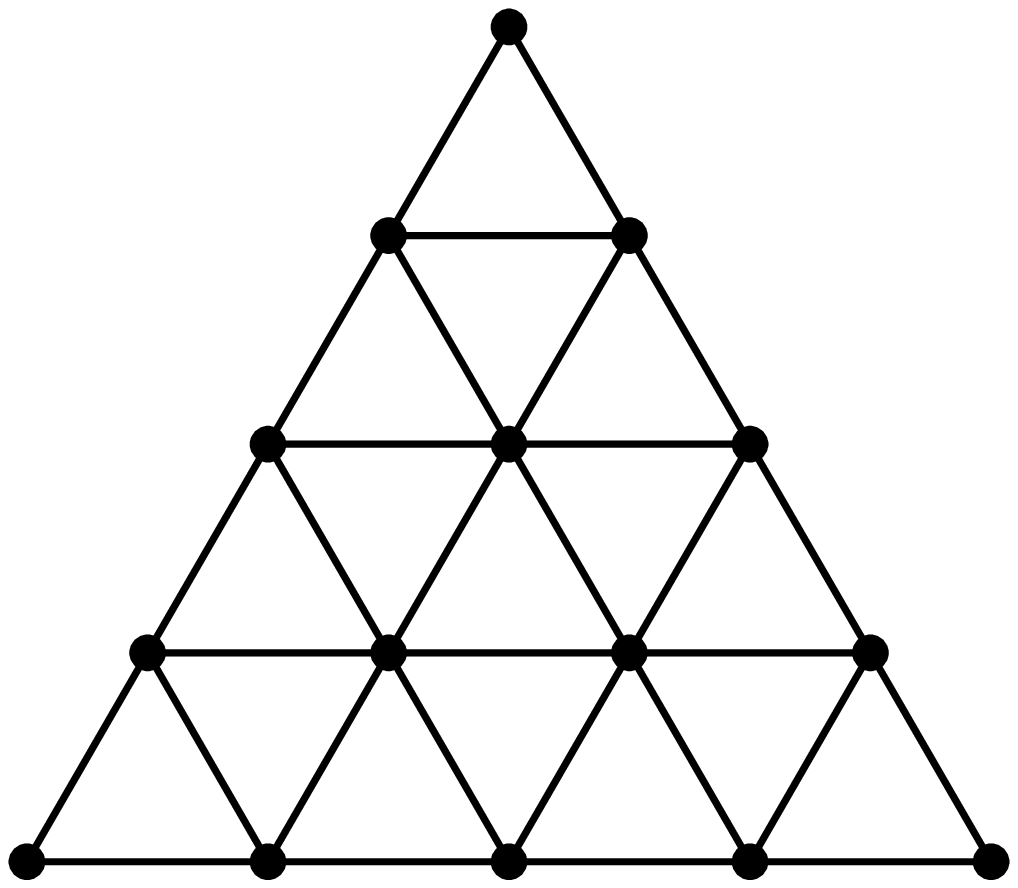}\label{fig:tria0}}
    \subfigure[T1]{\includegraphics[scale=0.3,clip]{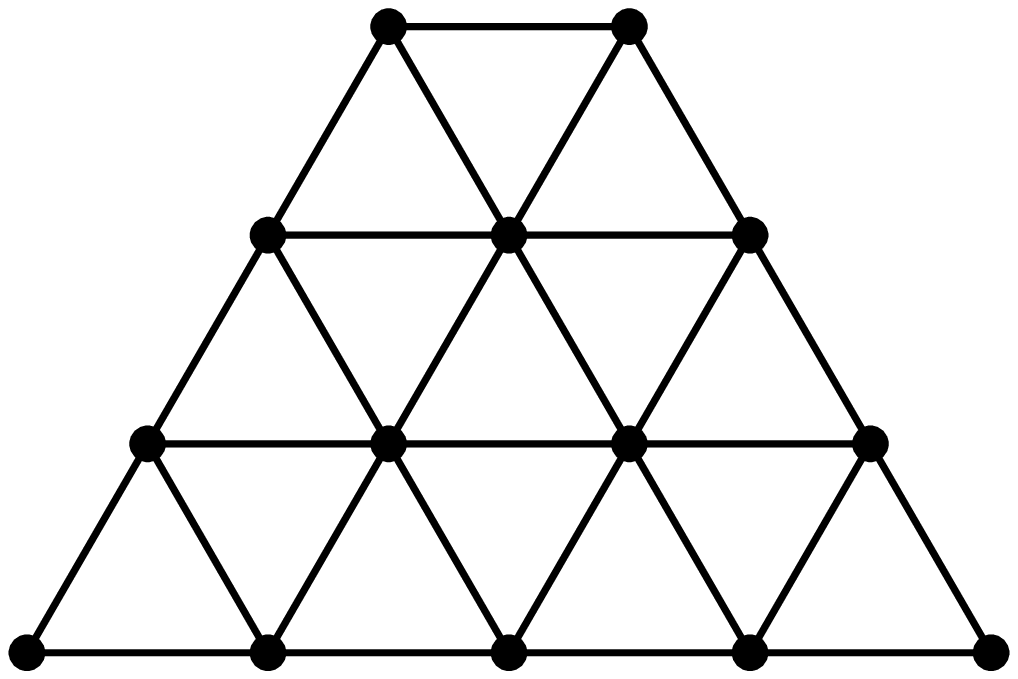}\label{fig:tria1}}
    \subfigure[T2]{\includegraphics[scale=0.3,clip]{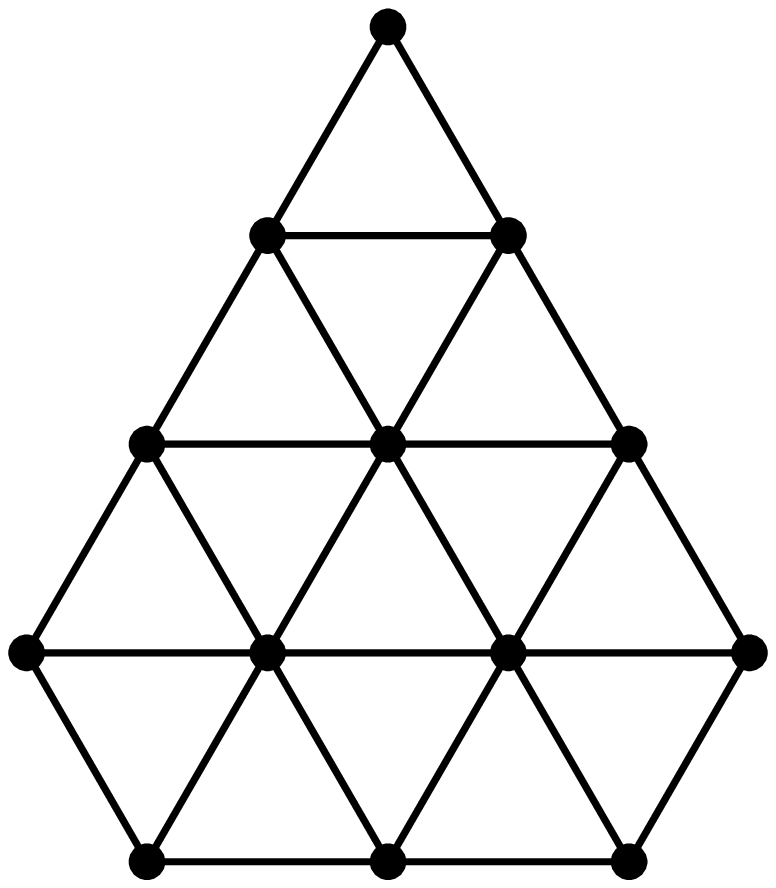}\label{fig:tria2}}
    \subfigure[T3]{\includegraphics[scale=0.3,clip]{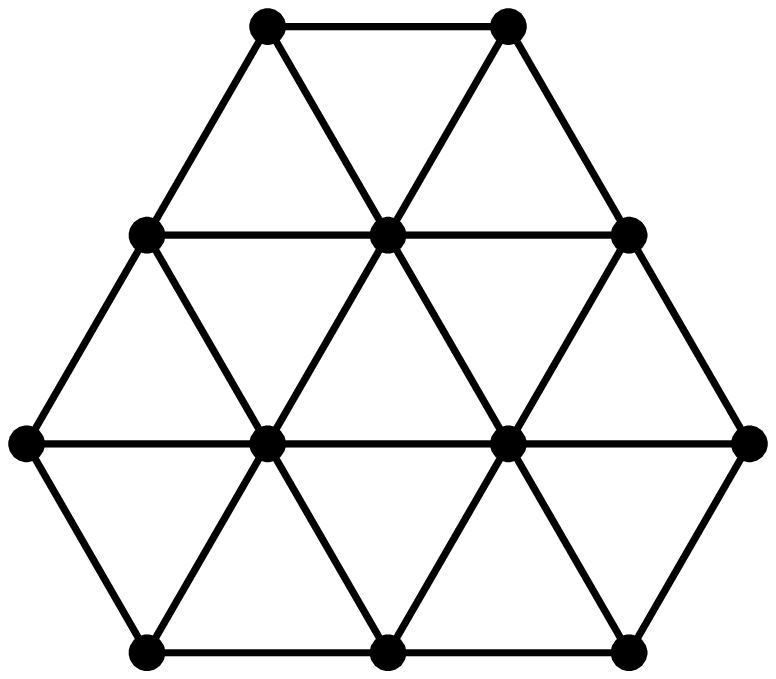}\label{fig:tria3}}\\
    \subfigure[L1]{\includegraphics[scale=0.3,clip]{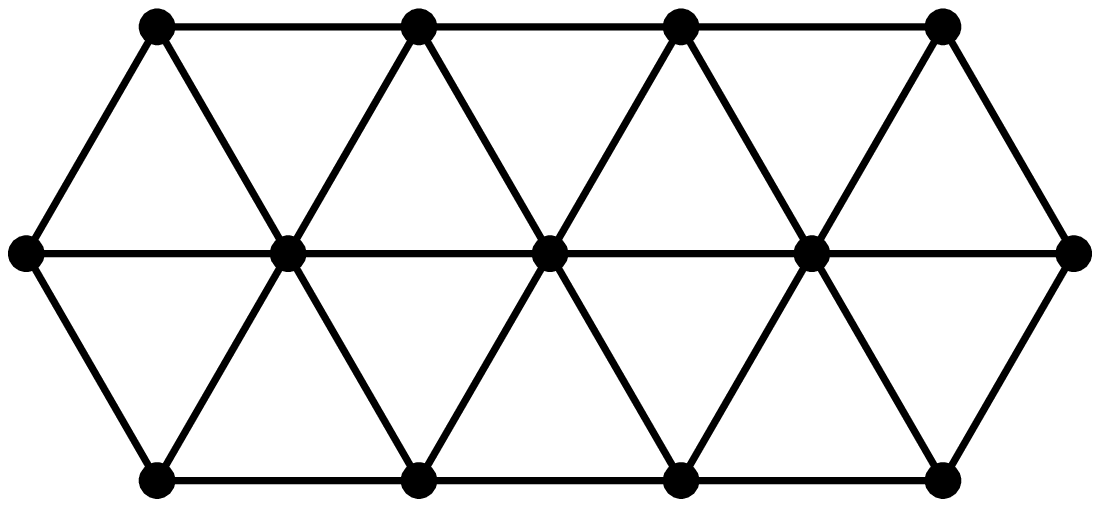}\label{fig:chain1}}
    \subfigure[L2]{\includegraphics[scale=0.3,clip]{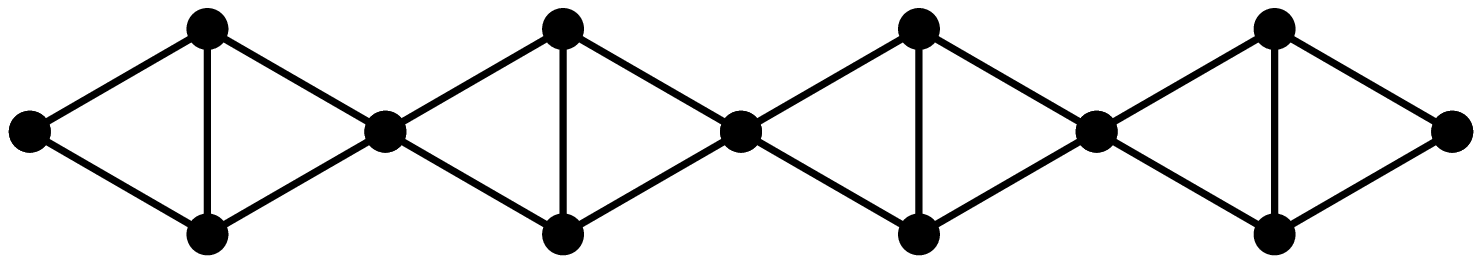}\label{fig:chain2}}
\caption{Various clusters derived from the rhombus with the side length $L=5$: (a) rhombus (R), (b) rhombus without the central spin (R1), (c) hexagon (H), (d) hexagon without the central spin (H1), (e) triangle (T), triangles with (f) one (T1), (g) two (T2) and (h) three (T3) vertices removed, ladder-like structures consisting of the spins along (i) the main diagonal (L1) and (j) the antidiagonal (L2) of the rhombus (a).}\label{fig:clusters}
\end{figure}

\hspace*{5mm} The spin-$s$ Ising model on a lattice with triangular geometry can be described by the Hamiltonian 
\begin{equation}
\label{Hamiltonian}
\mathcal H=-J\sum_{\langle i,j \rangle}s_{i}s_{j},
\end{equation}
where the spins on the $i$th lattice site are allowed to take $2s+1$ values: $s_{i}=-s,-s+1,\ldots,s-1,s$. The summation $\langle i,j \rangle$ runs over nearest-neighbor sites and $J<0$ is an antiferromagnetic exchange interaction parameter. The lattice is considered to consist of a finite number of spins arranged in clusters of various shapes, as shown in Fig.~\ref{fig:clusters}. For such relatively small clusters it is possible to fully explore the state space, exactly determine ground states and calculate some quantities of interest, such as the internal energy and the entropy. In order to understand the effects of different shapes and sizes of the finite clusters it is interesting to compare the obtained results with those for the thermodynamic limit, i.e., for an infinite lattice. The latter can be obtained from Monte Carlo (MC) simulation data and the residual (ground state) entropy can be determined by the thermodynamic integration method~\cite{kirk,bind}. 

\subsection{Exact enumeration} 
Considering the Hamiltonian~(\ref{Hamiltonian}) and for simplicity putting $J=-1$, a zero-temperature energy can be expressed as 
\begin{equation}
\label{Energy}
E=\sum_{\langle k,l \rangle}s_{k}s_{l},
\end{equation}
where the summation $\langle k,l \rangle$ runs over the nearest neighbors $s_{k}$ and $s_{l}$ on the considered cluster. Ground-state (GS) spin configurations ${\bf s_{\mathrm{GS}}}=\{s_1,\ldots,s_{N}\}$, where $N$ is a number of spins in the cluster, can be found as configurations\footnote{There are in total $(2s+1)^{N}$ configurations to be considered.} that minimize the energy functional $f({\bf s}) \equiv E$, i.e.:
\begin{eqnarray}
\label{GS} {\bf s_{\mathrm{GS}}} = \argmin_{{\bf s}} \, f({\bf s}).
\end{eqnarray}
We record the minimum energy per spin $E/N \equiv f({\bf s_{\mathrm{GS}}})/N$ and from the obtained number of GS configurations $W$, we calculate the entropy per spin $S/N=\ln W/N$ (we put the Boltzmann constant $k_B=1$).

\subsection{Monte Carlo simulation} 
For larger cluster sizes that cannot be approached by the exact enumeration we perform MC simulations employing the Metropolis dynamics and applying free boundary conditions. For thermal averaging we consider up to $2 \times 10^5$ MC sweeps after discarding another $20\%$ of the total number for thermalization. The simulations start from $\beta \equiv 1/T=0$ (i.e., $T = \infty$), using a random initial configuration, and proceed by $\beta$ being increased by the step $\Delta \beta$ and the simulation at $\beta+\Delta \beta$ starting from the final configuration obtained at $\beta$. We evaluate the internal energy $E=\langle \mathcal H \rangle$, which is subsequently used in the thermodynamic integration method~\cite{kirk,bind,roma} for calculation of the entropy as follows  
\begin{equation}
\label{TIM_E}
S(\beta)= N\ln (2s+1) + \beta E(\beta)- \int_{0}^{\beta}E(\beta')d(\beta').
\end{equation}
We note that using the above formula is more practical than the usual approach based on integrating the expression $dS(T)=(C/T)dT$, where $C$ is the specific heat. First of all, it is based on integration of the internal energy, which is directly measured in MC simulations. Another benefit is that it generally reduces errors resulting from the calculation of the specific heat $C$ from the energy $E$ and then from its numerical integration, particularly in case it shows a sharp peak. Finally, for each cluster size we perform $20$ simulation runs to suppress the sampling error. 
%


\section{Results}
\subsection{Ground state}
\subsubsection{Spin $s=1/2$}
\begin{figure}[t]
\centering
\subfigure[]{\includegraphics[scale=0.5,clip]{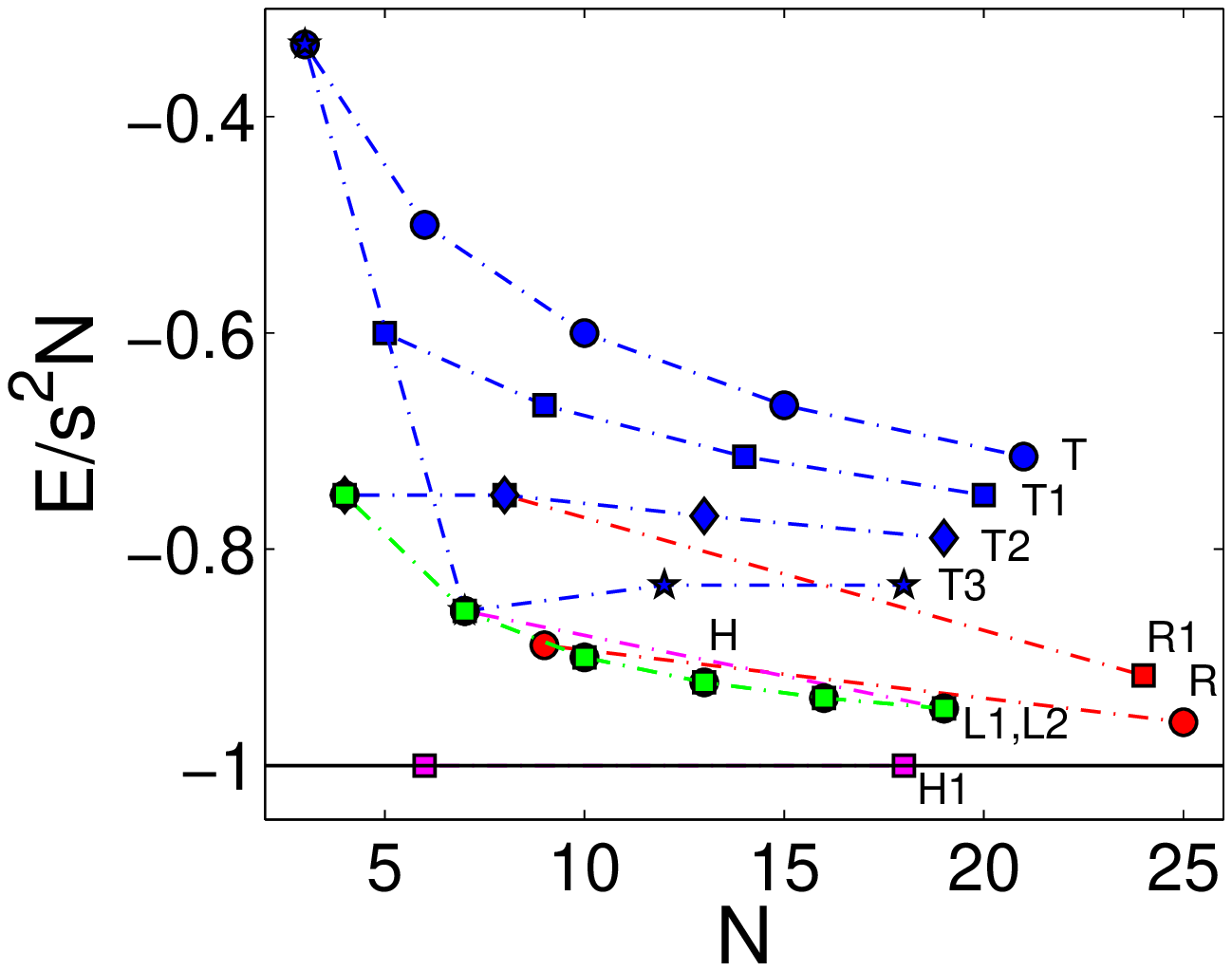}\label{fig:ene_s1_2}}
\subfigure[]{\includegraphics[scale=0.5,clip]{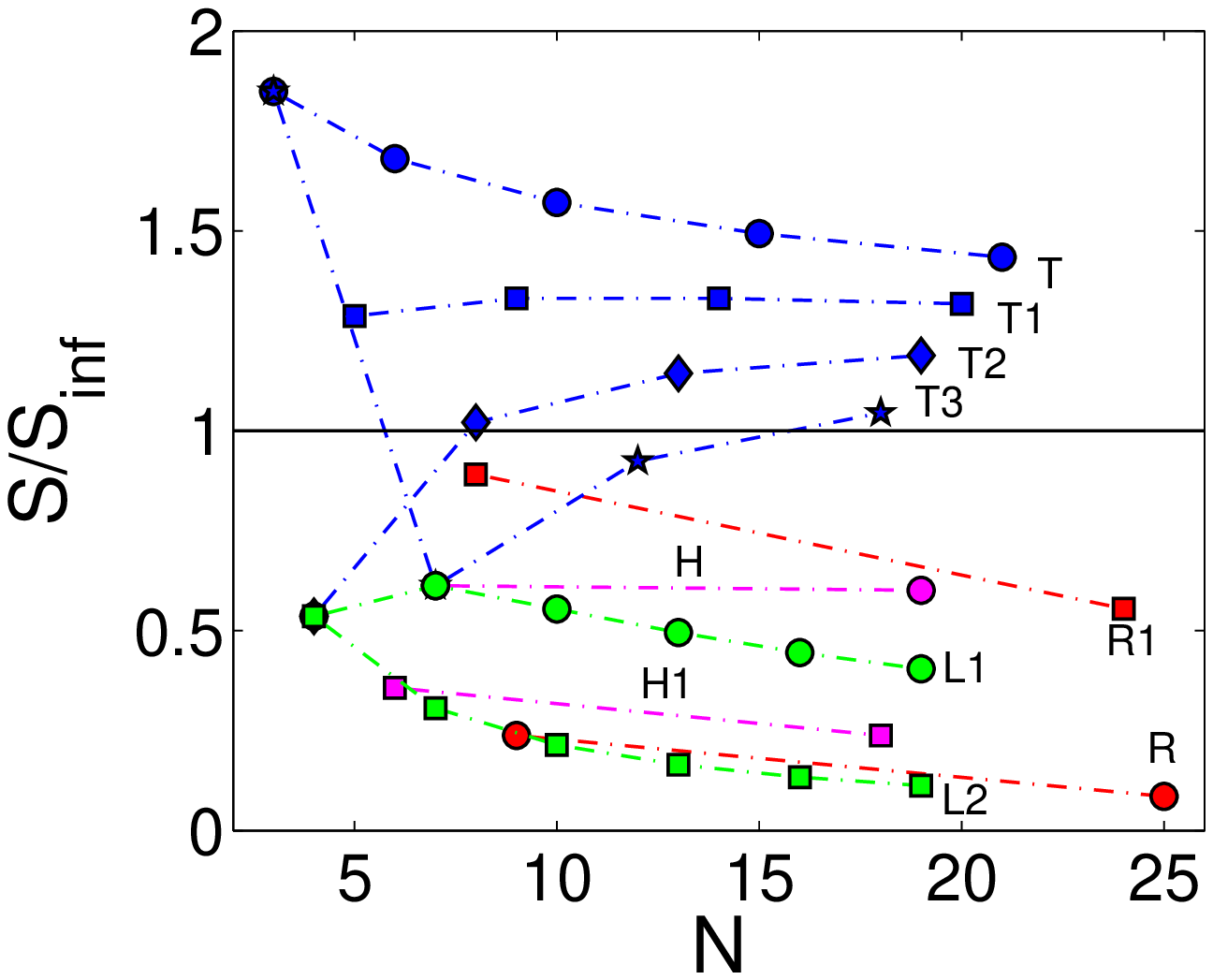}\label{fig:ent_s1_2}}
\caption{Reduced values of (a) the internal energy and (b) the entropy as functions of the cluster size for various cluster shapes and spin $s=1/2$. $S_{inf}$ represents the entropy of the infinite system.}\label{fig:ene_ent_s1_2}
\end{figure}

The ground-state configurations of finite clusters, their energies and degeneracies are expected to depend on the shapes and sizes of the clusters~\cite{milla1,milla2,milla3}. The exact results presented in Table~\ref{tab:entropy_s1_2} demonstrate how the respective quantities for different cluster shapes change with the increasing cluster size for spin $s=1/2$. For better visual presentation, as well as relation of the obtained values with that expected in the infinite lattice, we show the following plots: In Fig.~\ref{fig:ene_s1_2} we plot the reduced internal energies per spin $E/s^2N$ and in Fig.~\ref{fig:ent_s1_2} the reduced entropy $S/S_{inf}$, where $S_{inf} = 0.3231$ represents the exact Wannier's value of the infinite system~\cite{wann2}, as functions of the spin numbers $N$. Thus, in the thermodynamic limit of $N \to \infty$ the quantities $E/s^2N$ are expected to approach the value of $-1$ and $S/S_{inf}$ the value of $1$~\cite{wann}. The respective values for the original rhombic (R), hexagonal (H) and triangular (T) shapes correspond to those presented in Ref.~\cite{milla1}\footnote{Actually, the values of $S/S_{inf}$ slightly deviate, since in Ref.~\cite{milla1} incorrect value of $S_{inf} = 0.3383$ was used.} for the chosen cluster sizes. With the increasing size, all the energies seem to approach the infinite limit value of $-1$ and the entropies of T and H shapes the value of $1$. On the other hand, the entropy of the rhombic clusters R vanishes as the cluster size increases.\\
\hspace*{5mm}  Now, let us investigate how this behavior can change upon minor manipulations with the cluster shapes, such as removal of one, two or three spins from the cluster. From Fig.~\ref{fig:ene_s1_2} it is apparent that the gradual removal of the spins from the vertices of the triangular cluster causes decrease in the energy, which is more rapid in smaller clusters. This eventually changes the monotonically decreasing character of the energy vs. cluster size function (see e.g., T) to the curve showing an initial increase for smaller clusters followed by a gradual decrease to the infinite-limit value (T3). The effects of one spin removal from the center of rhombus (R1) and hexagon (H1) on the energy are just opposite. In the former case the energy is increased while in the latter decreased. In fact, the energy of H1 cluster retains the minimum value of $-1$, regardless of the cluster size, which can be understood as follows. The non-frustrated elementary H1 hexagon consisting of $N_0=6$ spins has the minimum energy if the spins alternate around the circumference and its energy equals to $E_0/s^2=-N_0$. By adding another outer hexagonal shell consisting of $N_1=12$ spins with the alternating values the minimum energy condition is preserved and the total energy will then change by $E_1/s^2=-N_1$, since the intershell contributions cancel due to equal number of satisfied and unsatisfied bonds. This scenario will repeat each time another shell is added. Thus, the total reduced energy of the H1 cluster of the size $N$ equals to $E/s^2=-N$ and hence the energy per spin equals to $-1$. The remaining ladder-like shapes L1 and L2 have the same energies, which relatively quickly approach the value of $-1$. \\
\hspace*{5mm} In Fig.~\ref{fig:ent_s1_2} we present effects of the cluster shape manipulation on the reduced entropy. Similar to the energy behavior, the entropy of the triangle-based shapes changes its monotonically decreasing character upon gradual spin removal. However, here the entropy as a function of the number of spins $N$ becomes non-monotonic already when one vertex is removed (T1), showing an initial increase followed by a gradual decrease for larger $N$. For T2 and T3 the entropy of the smallest clusters is even less than the infinite limit value. Nevertheless, with the increasing cluster size it exceeds $S_{inf}$ before it starts asymptotically approaching it again from above. Of course, here we assume that the infinite limit is not affected by such a spin manipulation and that $S=S_{inf}$ for all the triangle-based shapes. The entropies of the shapes R and H show qualitatively similar changes as the energy when one spin is removed. Namely, the former increases while the latter decreases. For the considered cluster sizes the changes are rather dramatic, particularly the entropy change between R and R1. Interesting is a noticeable entropy reduction when H1 is increased from one-shell ($L=3$) to two-shell ($L=5$) structure. As can be seen in Table~\ref{tab:entropy_s1_2}, the latter has only four degenerate states in which the two shells are formed by alternating spin values and both can be globally inverted without energy change. Nevertheless, by adding further shells there will be a number of other energetically equivalent configurations and therefore we expect the entropy to increase and eventually approach $S_{inf}$ for $N \to \infty$. The shape L2 is like the rhombic shape R non-degenerate (apart from global spin inversion) and thus the quantity $S/S_{inf}$ quickly approaches zero. On the other hand, degeneracy of the L1 shape increases when $L$ is incremented by one (which means the increase of $N$ by three) but only by a constant number of states (two for $s=1/2$) and, hence, in this case $S/N=\ln(2L-2)/(3L-2)$ also vanishes in the thermodynamic limit.
 
\begin{table}[]
\caption{Ground-state energy per spin $E/N$, number of degenerate states $W$ and entropy per spin $S/N$ for different spin clusters with the size $N$ and the spin value $s=1/2$.}
\label{tab:entropy_s1_2}
\centering
\scalebox{0.95}{
\renewcommand{\arraystretch}{0.99}
\begin{tabular}{cc|cccccc}
\hline
 &  &  \multicolumn{6}{c}{$L$} \\\hline
Cluster &       &     2     &     3     &     4     &     5     &     6     &     7      \\
\hline

R       & $N$   & -         &     9     &     -     &     25    &     -     &     - \\ 
        & $E/N$ & -         &   -0.2222 &     -     &   -0.2400 &     -     &     - \\
        & $W$   & -         &     2     &     -     &     2     &     -     &     - \\
        & $S/N$ & -         &    0.0770 &     -     &    0.0277 &     -     &     - \\
R1      & $N$   & -         &     8     &     -     &     24    &     -     &     - \\ 
        & $E/N$ & -         &   -0.1875 &     -     &   -0.2292 &     -     &     - \\
        & $W$   & -         &      10   &     -     &     74    &     -     &     - \\
        & $S/N$ & -         &    0.2878 &     -     &    0.1793 &     -     &     - \\       
H       & $N$   & -         &     7     &     -     &     19    &     -     &     - \\ 
        & $E/N$ & -         &   -0.2143 &     -     &   -0.2368 &     -     &     - \\
        & $W$   & -         &      4    &     -     &     40    &     -     &     - \\
        & $S/N$ & -         &    0.1980 &     -     &    0.1942 &     -     &     - \\
H1      & $N$   & -         &      6    &     -     &      18   &     -     &     - \\ 
        & $E/N$ & -         &   -0.2500 &     -     &   -0.2500 &     -     &     - \\
        & $W$   & -         &      2    &     -     &      4    &     -     &     - \\
        & $S/N$ & -         &    0.1155 &     -     &    0.0770 &     -     &     - \\
T       & $N$   & 3         &         6 &        10 &        15 &        21 & - \\ 
        & $E/N$ & -0.0833   &   -0.1250 &   -0.1500 &   -0.1667 &   -0.1786 & - \\
        & $W$   & 6        &       26   &       160 &     1386  &    16814  & - \\
        & $S/N$ & 0.5973    &    0.5430 &    0.5075 &    0.4823 &    0.4633 & - \\
T1      & $N$   & -         &     5     &         9 &        14 &        20 & - \\ 
        & $E/N$ & -         &   -0.1500 &   -0.1667 &   -0.1786 &   -0.1875 & - \\
        & $W$   & -         &         8 &        48 &     412   &     4980  & - \\
        & $S/N$ & -         &    0.4159 &    0.4301 &    0.4301 &    0.4257 & - \\
T2      & $N$   & -         &     4     &         8 &        13 &        19 & - \\ 
        & $E/N$ & -         &   -0.1875 &   -0.1875 &  -0.1923  &   -0.1974 & - \\
        & $W$   & -         &      2    &      14   &   122     &     1474  & - \\
        & $S/N$ & -         &    0.1733 &   0.3299  &    0.3695 &    0.3840 & - \\             
T3      & $N$   & -         &     3     &         7 &        12 &        18 & - \\ 
        & $E/N$ & -         &  -0.0833  &   -0.2143 &   -0.2083 &   -0.2083 & - \\
        & $W$   & -         &     6     &    4      &    36     &    436    & - \\
        & $S/N$ & -         &    0.5973 &    0.1980 &    0.2986 &   0.3376  & - \\
L1      & $N$   &     4     &     7     &     10    &     13   &     16    &   19 \\ 
        & $E/N$ &  -0.1875  &   -0.2143 &   -0.2250 &  -0.2308 &   -0.2344 &   -0.2368 \\
        & $W$   &     2     &     4     &     6     &     8    &     10    &   12 \\
        & $S/N$ & 0.1733    &    0.1980 &    0.1792 &   0.1600 &    0.1439 &   0.1308 \\        
L2      & $N$   &     4     &     7     &     10    &     13   &     16    &   19 \\ 
        & $E/N$ &  -0.1875  &   -0.2143 &   -0.2250 &  -0.2308 &   -0.2344 &  -0.2368 \\
        & $W$   &     2     &     2     &     2     &      2   &    2      &    2 \\
        & $S/N$ &   0.1733  &    0.0990 &    0.0693 &   0.0533 &   0.0433  &   0.0365 \\        
\end{tabular}}
\end{table}

\subsubsection{Spin $s>1/2$}

\begin{figure}[t]
\centering
    \subfigure[s=1]{\includegraphics[scale=0.5,clip]{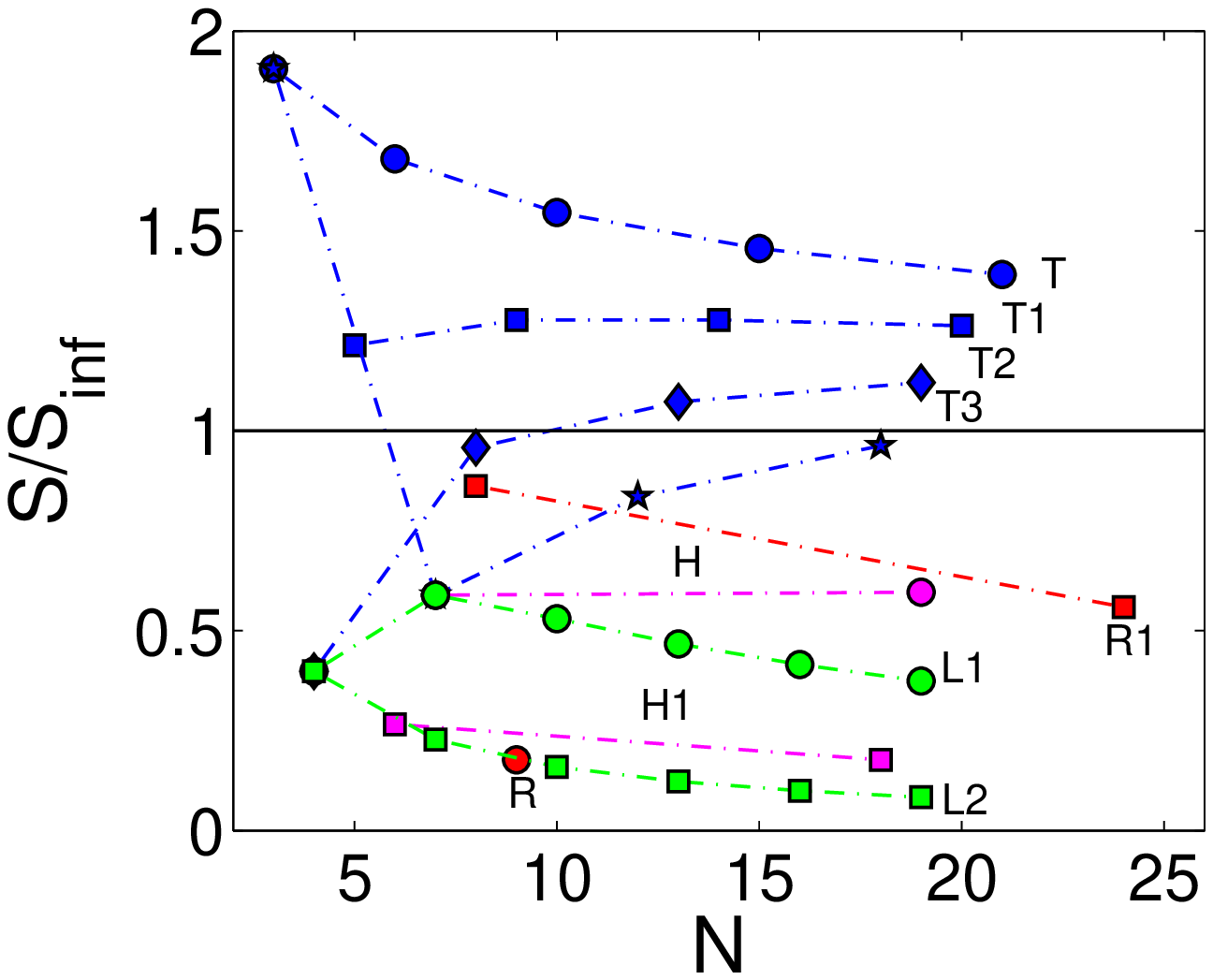}\label{fig:ent_s1}}
    \subfigure[s=3/2]{\includegraphics[scale=0.5,clip]{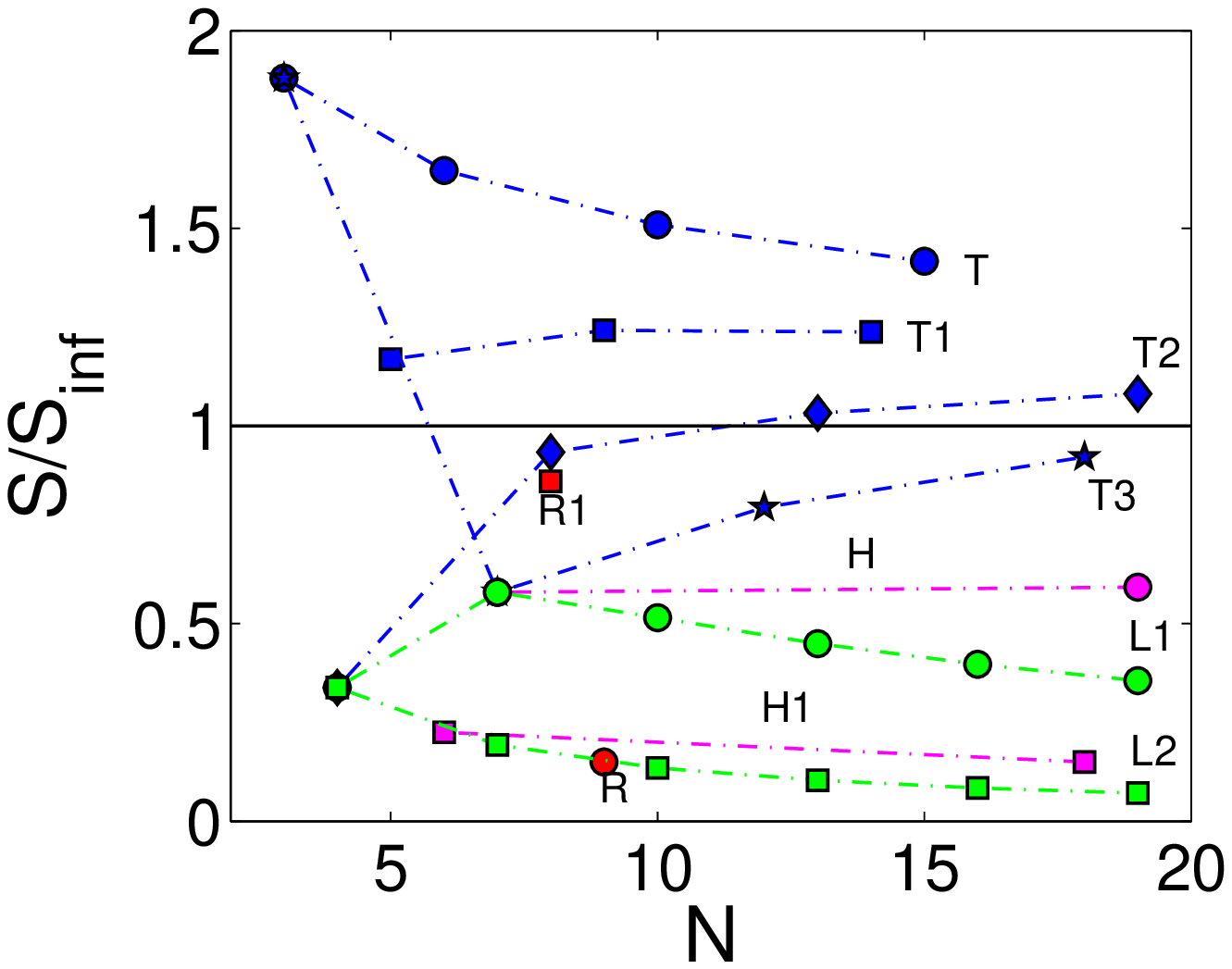}\label{fig:ent_s3_2}} 
\caption{Reduced entropy as a function of the cluster size for various cluster shapes and spins (a) $s=1$ and (b) $s=3/2$. }\label{fig:ent_s_all}
\end{figure}

\begin{figure}[t]
\centering
    \subfigure[T,T1,T2,T3]{\includegraphics[scale=0.5,clip]{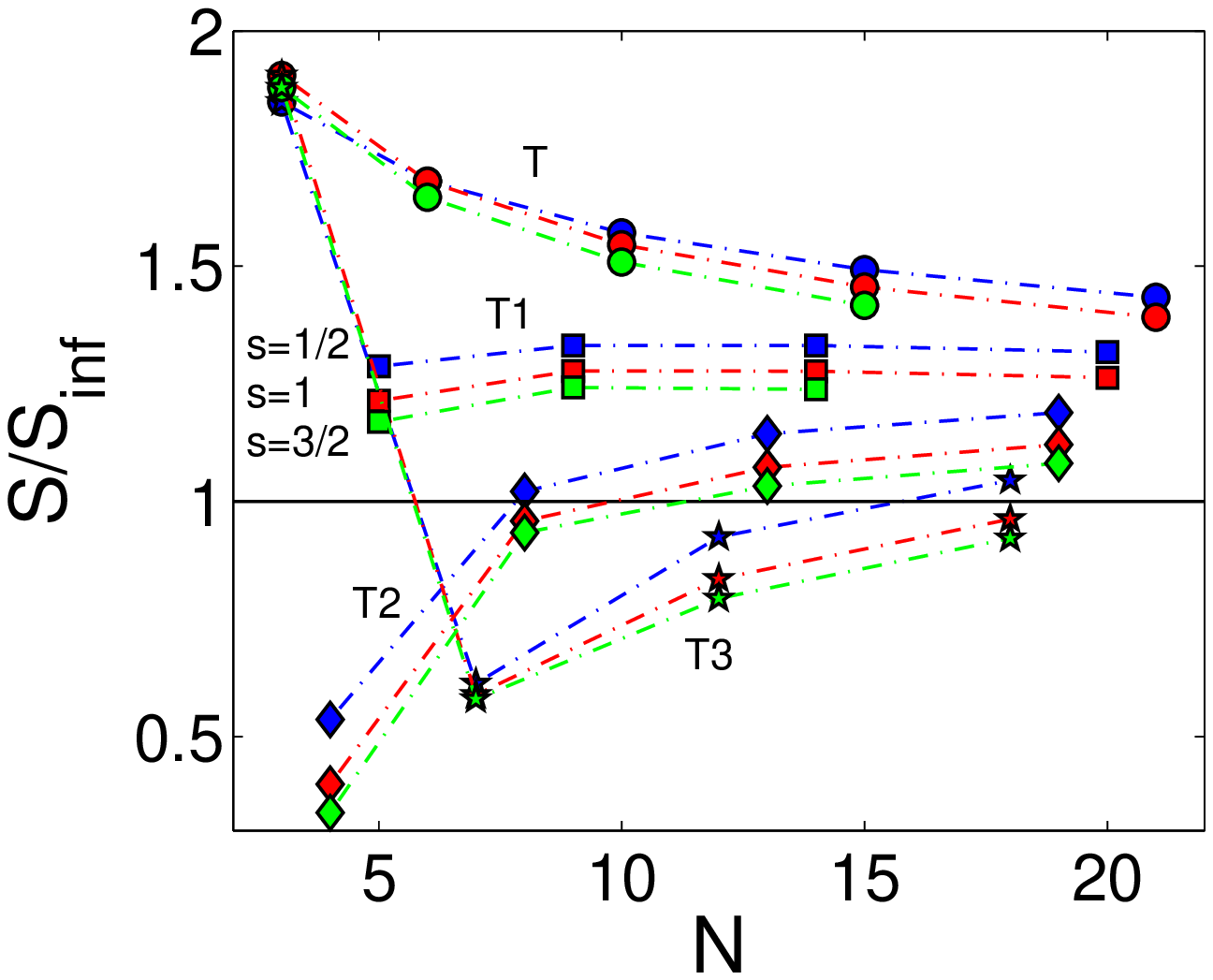}\label{fig:ent_s_all_tria}}
    \subfigure[R,R1]{\includegraphics[scale=0.5,clip]{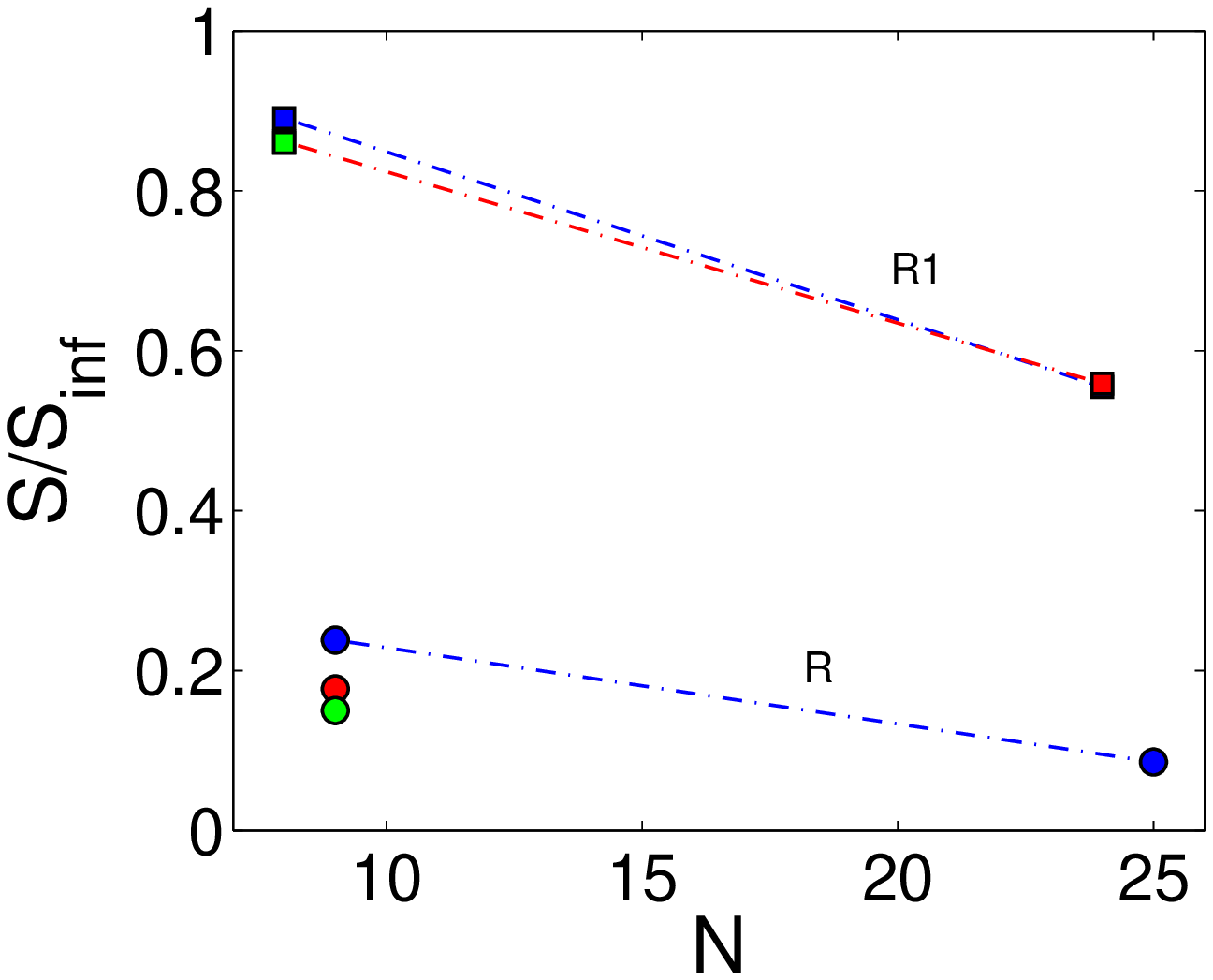}\label{fig:ent_s_all_rhomb}}
    \subfigure[H,H1]{\includegraphics[scale=0.5,clip]{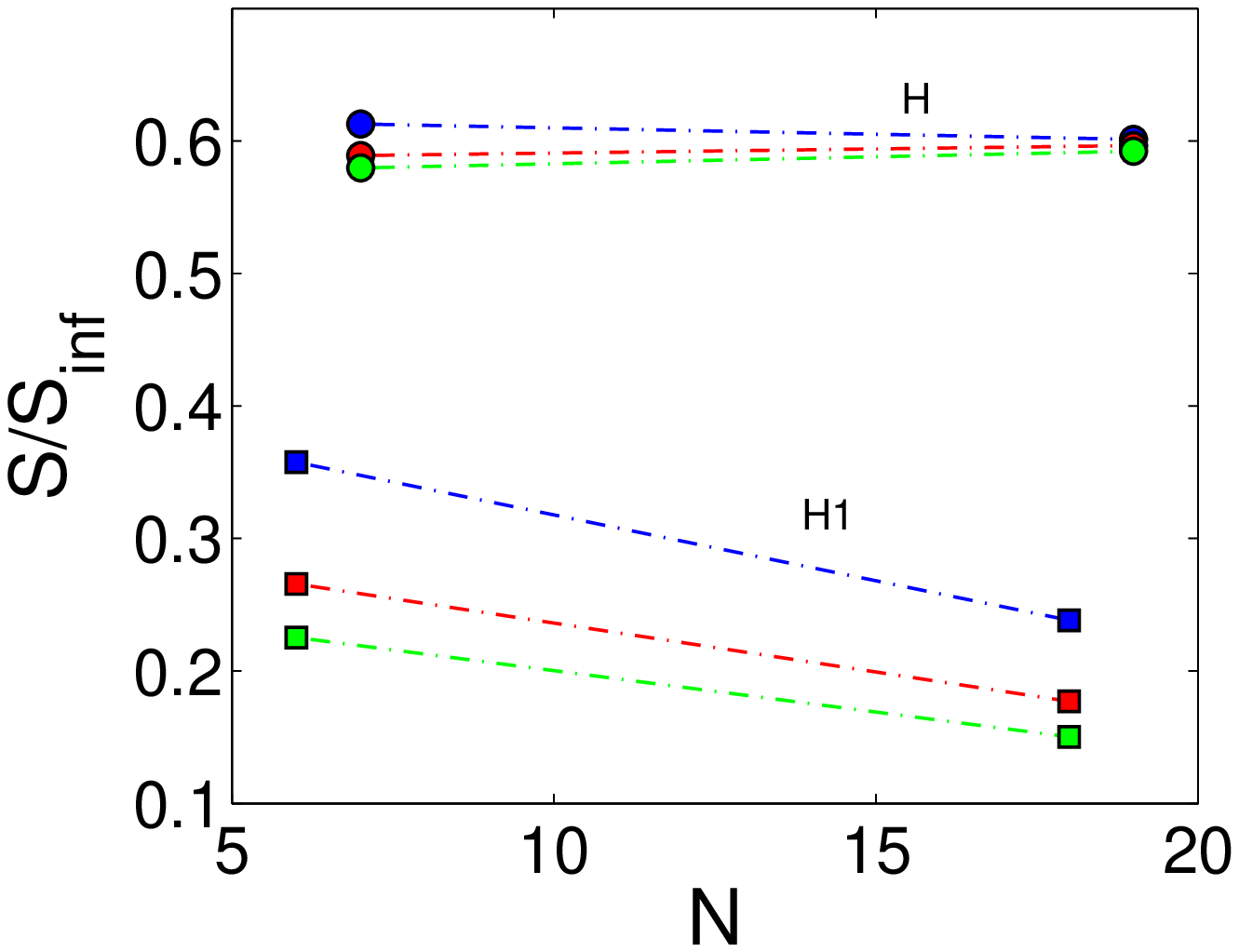}\label{fig:ent_s_all_hexa}} 
    \subfigure[L1,L2]{\includegraphics[scale=0.5,clip]{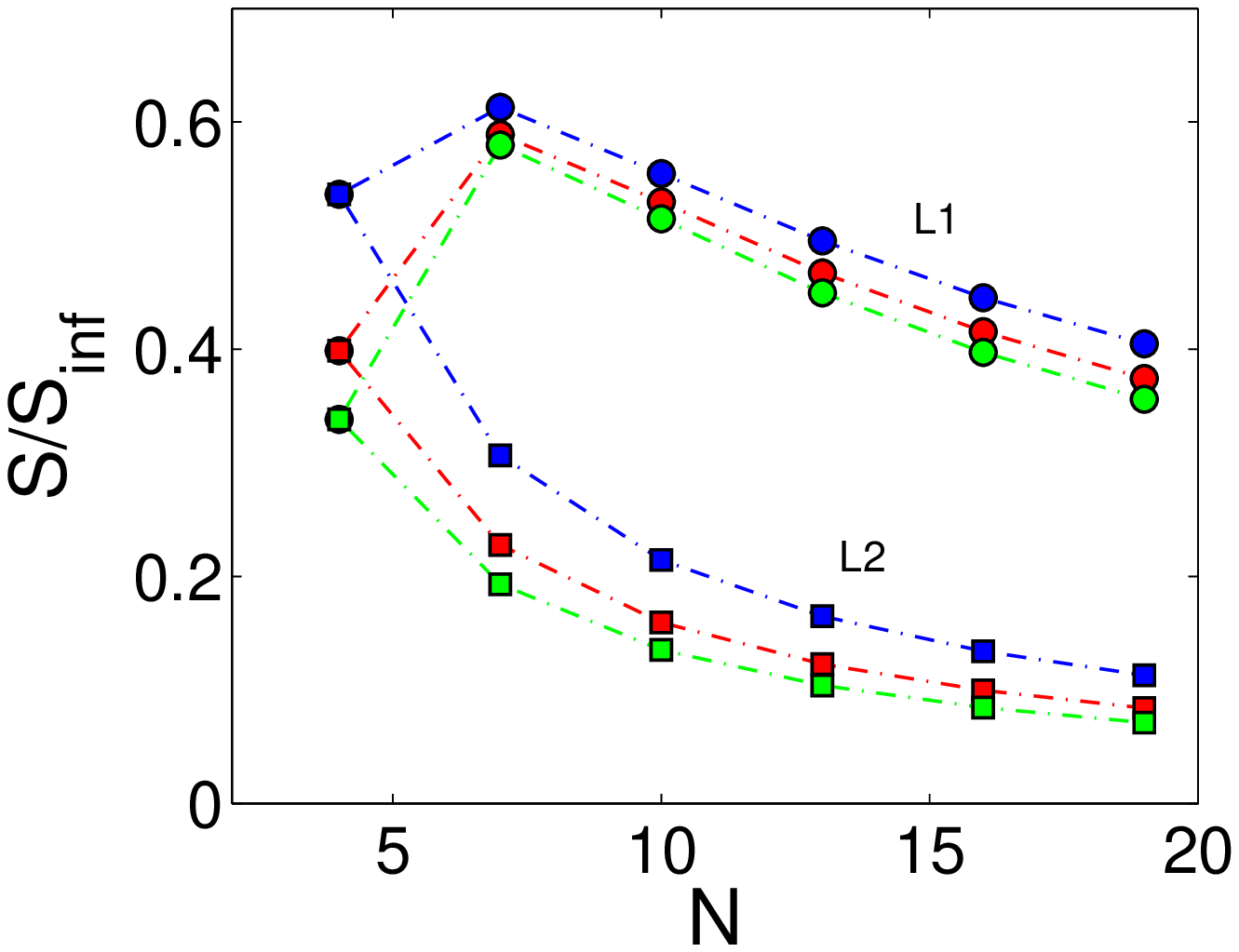}\label{fig:ent_s_all_chain}}
\caption{Reduced entropy as a function of the cluster size for spins $s=1/2$ (blue), $s=1$ (red) and $s=3/2$ (green), and various cluster shapes. }\label{fig:ent_s_all_comp}
\end{figure}

In the following we present results for the clusters with the spin $s>1/2$. Numerical values, similar to those shown in Table~\ref{tab:entropy_s1_2} for $s=1/2$, are presented in Table~\ref{tab:entropy_s1} for $s=1$ and Table~\ref{tab:entropy_s3_2} for $s=3/2$. We note that some values for larger clusters are missing due to difficulties to scan the exponentially increasing configuration space for larger degrees of freedom in to order to perform the exact enumeration in reasonable CPU times. By inspecting the reduced energy $E/s^2N$, it is easy to verify that for all the cluster shapes and sizes the values are {\it equal} and thus do not depend on the spin value. This finding implies that for a given cluster shape and size all the degenerate states involve configurations of only extremal spin values ($s_i=\pm s$) with the same configuration of signs. Looking at the entropy, in general, the increased spin value results in larger ground-state degeneracy and hence larger residual entropy~\cite{zuko2}. However, this is not the case for R and L2 shapes, which remain non-degenerate regardless of the spin value. For better comparison with the spin $s=1/2$ cases, it is convenient to display in Fig.~\ref{fig:ent_s_all} the reduced quantities, such as those shown in Fig.~\ref{fig:ent_s1_2} for $s=1/2$. The respective values for the infinite systems, $S_{inf}=0.4347$ for $s=1$ and $S_{inf}=0.5126$ for $s=3/2$, have been determined by MC simulations~\cite{zuko2}, applying the thermodynamic integration method~\cite{roma}. Although the entropy $S$ changes, as presented in Tables~\ref{tab:entropy_s1} and~\ref{tab:entropy_s3_2}, the variation of the relative values $S/S_{inf}$ with the spin value can barely be observed. Nevertheless, a closer inspection reveals some spin dependence which is better visualized in Fig.~\ref{fig:ent_s_all_comp} and which shows separately the cases of (a) triangle- (b) rhombus- (c) hexagon- and (d) ladder-based clusters for spin 1/2 (blue), 1 (red) and 3/2 (green). For virtually all the cases the spin-dependence of the reduced entropy $S/S_{inf}$ can be characterized as decreasing. There are two exceptions, though, in which $S/S_{inf}$ achieves the smallest value for $s=1/2$: the elementary triangle T with three vertices and R1 shape with 24 spins.

\begin{table}[]
\caption{The same as in Table~\ref{tab:entropy_s1_2} for the spin value $s=1$.}
\label{tab:entropy_s1}
\centering
\scalebox{0.95}{
\renewcommand{\arraystretch}{0.99}
\begin{tabular}{cc|cccccc}
\hline
 &  &  \multicolumn{6}{c}{$L$} \\\hline
Cluster &       &     2     &     3     &     4     &     5     &     6     &     7      \\
\hline

R       & $N$   & -         &     9     &     -     &      -    &     -     &     - \\ 
        & $E/N$ & -         &   -0.8889 &     -     &      -    &     -     &     - \\
        & $W$   & -         &     2     &     -     &      -    &     -     &     - \\
        & $S/N$ & -         &    0.0770 &     -     &      -    &     -     &     - \\
R1      & $N$   & -         &     8     &     -     &     24    &     -     &     - \\ 
        & $E/N$ & -         &   -0.7500 &     -     & -0.9167   &     -     &     - \\
        & $W$   & -         &      20   &     -     &    340    &     -     &     - \\
        & $S/N$ & -         &    0.3745 &     -     & 0.2429    &     -     &     - \\        
H       & $N$   & -         &     7     &     -     &     19    &     -     &     - \\ 
        & $E/N$ & -         &   -0.8571 &     -     &   -0.9474 &     -     &     - \\
        & $W$   & -         &      6    &     -     &     138   &     -     &     - \\
        & $S/N$ & -         &    0.2560 &     -     &    0.2593 &     -     &     - \\
H1      & $N$   & -         &      6    &     -     &     18    &     -     &     - \\ 
        & $E/N$ & -         &   -1.0000 &     -     &   -1.0000 &     -     &     - \\
        & $W$   & -         &      2    &     -     &      4    &     -     &     - \\
        & $S/N$ & -         &    0.1155 &     -     &    0.0770 &     -     &     - \\
T       & $N$   & 3         &         6 &        10 &        15 &        21 & - \\ 
        & $E/N$ & -0.3333   &      -0.5 &      -0.6 &   -0.6667 &   -0.7143 & - \\
        & $W$   & 12        &      80   &       828 &     13242 &    327206 & - \\
        & $S/N$ & 0.8283    &    0.7303 &    0.6719 &    0.6327 &    0.6047 & - \\
T1      & $N$   & -         &     5     &         9 &        14 &        20 & - \\ 
        & $E/N$ & -         &      -0.6 &   -0.6667 &   -0.7143 &   -0.7500 & - \\
        & $W$   & -         &        14 &       148 &     2368  &     58550 & - \\
        & $S/N$ & -         &    0.5278 &    0.5552 &    0.5550 &    0.5489 & - \\
T2      & $N$   & -         &     4     &         8 &        13 &        19 & - \\ 
        & $E/N$ & -         &     -0.75 &    -0.75  &  -0.7692  &   -0.7895 & - \\
        & $W$   & -         &      2    &      28   &   428     &     10456 & - \\
        & $S/N$ & -         &    0.1733 &   0.4165  &    0.4661 &    0.4871 & - \\             
T3      & $N$   & -         &     3     &         7 &        12 &        18 & - \\ 
        & $E/N$ & -         &  -0.3333  &   -0.8571 &   -0.8333 &   -0.8333 & - \\
        & $W$   & -         &     12    &    6      &    78     &    1864   & - \\
        & $S/N$ & -         &    0.8283 &    0.2560 &    0.3631 &   0.4184  & - \\
L1      & $N$   &     4     &     7     &     10     &     13   &     16    &   19 \\ 
        & $E/N$ &  -0.75    &   -0.8571 &    -0.9    &  -0.9231 &   -0.9375 &   -0.9474 \\
        & $W$   &     2     &     6     &     10     &     14   &     18    &   22 \\
        & $S/N$ & 0.1733    &    0.2560 &     0.2303 &   0.2030 &    0.1806 &   0.1627 \\        
L2      & $N$   &     4     &     7     &     10     &     13   &     16    &   19 \\ 
        & $E/N$ &  -0.75    &   -0.8571 &   -0.9000  &  -0.9231 &   -0.9375 &  -0.9474 \\
        & $W$   &     2     &     2     &     2      &      2   &    2      &    2 \\
        & $S/N$ &   0.1733  &    0.0990 &    0.0693  &   0.0533 &   0.0433  &   0.0365 \\        
\end{tabular}}
\end{table}

\begin{table}[t!]
\caption{The same as in Table~\ref{tab:entropy_s1_2} for the spin value $s=3/2$.}
\label{tab:entropy_s3_2}
\centering
\renewcommand{\arraystretch}{0.98}
\begin{tabular}{cc|cccccc}
\hline
 &  &  \multicolumn{6}{c}{$L$} \\\hline
Cluster &       &     2     &     3     &     4     &     5     &     6     &     7      \\
\hline

R       & $N$   &   -       &     9     &     -     &     -     &     -     &     - \\  
        & $E/N$ &   -       &    -2     &     -     &     -     &     -     &     - \\ 
        & $W$   &   -       &     2     &     -     &     -     &     -     &     - \\ 
        & $S/N$ & -         &  0.0770   &     -     &     -     &     -     &     - \\ 
        
R1      & $N$   & -         &     8     &     -     &     -     &     -     &     - \\  
        & $E/N$ & -         &   -1.6875 &     -     &     -     &     -     &     - \\ 
        & $W$   & -         &     34    &     -     &     -     &     -     &     - \\ 
        & $S/N$ & -         &    0.4408 &     -     &     -     &     -     &     - \\  
        
H       & $N$   & -         &     7     &     -     &     19    &     -     &     - \\ 
        & $E/N$ & -         &  -1.9286  &     -     &    -2.131 &     -     &     - \\
        & $W$   & -         &     8     &     -     &     320   &     -     &     - \\
        & $S/N$ & -         &   0.2971  &     -     &    0.3036 &     -     &     - \\

H1      & $N$   & -         &     6     &     -     &     18     &     -     &     - \\ 
        & $E/N$ & -         &   -2.2500 &     -     &    -2.2500 &     -     &     - \\
        & $W$   & -         &     2     &     -     &     4   &     -     &     - \\
        & $S/N$ & -         &    0.1155 &     -     &     0.0770 &     -     &     - \\        
        
T       & $N$   &     3     &     6     &    10     &    15     &    -     & - \\ 
        & $E/N$ &  -0.7500  &   -1.1250 &   -1.3500 &   -1.5000 &    -     & - \\
        & $W$   &     18    &     158   &     2280  &     53814 &    -     & - \\
        & $S/N$ &    0.9635 &    0.8438 &    0.7732 &    0.7262 &    -     & - \\
        
T1      & $N$   & -         &     5     &     9     &    14     &    -     & - \\ 
        & $E/N$ & -         &   -1.3500 &   -1.5000 &   -1.6071 &    -     & - \\
        & $W$   & -         &     20    &      308  &     7212  &    -     & - \\
        & $S/N$ & -         &   0.5991  &    0.6367 &    0.6345 &    -     & - \\
        
T2      & $N$   & -         &     4     &         8 &    13     &        19 & - \\ 
        & $E/N$ & -         &   -1.6875 &   -1.6875 &   -1.7308 &   -1.7763 & - \\
        & $W$   & -         &     2     &    46     &     974   &     37362 & - \\
        & $S/N$ & -         &   0.1733  &    0.4786 &    0.5293 &    0.5541 & - \\             
        
T3      & $N$   & -         &     3     &     7     &        12 &        18 & - \\ 
        & $E/N$ & -         &   -0.7500 &   -1.9286 &  -1.8750  &    -1.8750 & - \\
        & $W$   & -         &     18    &     8     &    132    &     4924  & - \\
        & $S/N$ & -         &    0.9635 &    0.2971 &    0.4069 &     0.4723 & - \\
        
L1      & $N$   &     4     &     7     &     10     &     13   &     16    &   19 \\ 
        & $E/N$ &   -1.6875 &   -1.9286 &    -2.0250 &  -2.0769 &   -2.1094 &   -2.1316\\
        & $W$   &     2     &      8    &      14    &    20    &      26   &   32 \\
        & $S/N$ &    0.1733 &    0.2971 &    0.2639  &   0.2304 &    0.2036 &   0.1824\\        
        
L2      & $N$   &     4     &     7     &     10     &     13   &     16    &   19 \\ 
        & $E/N$ &   -1.6875 &   -1.9286 &   -2.0250  &  -2.0769 &   -2.1094 &  -2.1316 \\
        & $W$   &     2     &     2     &      2     &     2    &      2    &    2 \\
        & $S/N$ &   0.1733  &   0.0990  &    0.0693  &  0.0533  &    0.0433 &   0.0365 \\        
\end{tabular}
\end{table}



\subsection{Finite-temperature behavior}
\begin{figure}[t]
\centering
    \subfigure[]{\includegraphics[scale=0.5,clip]{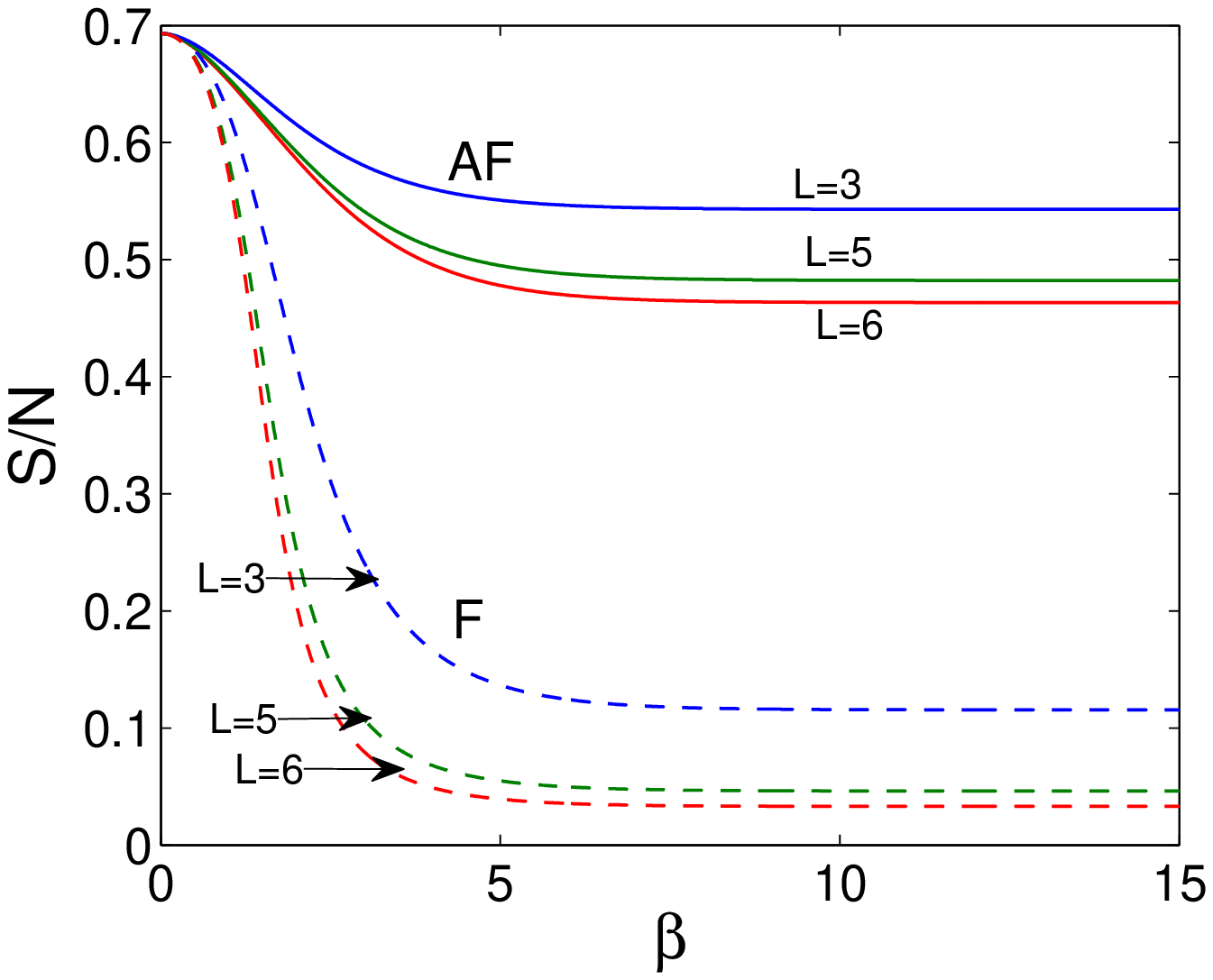}\label{fig:ent-fin_T_T0}}
    \subfigure[]{\includegraphics[scale=0.5,clip]{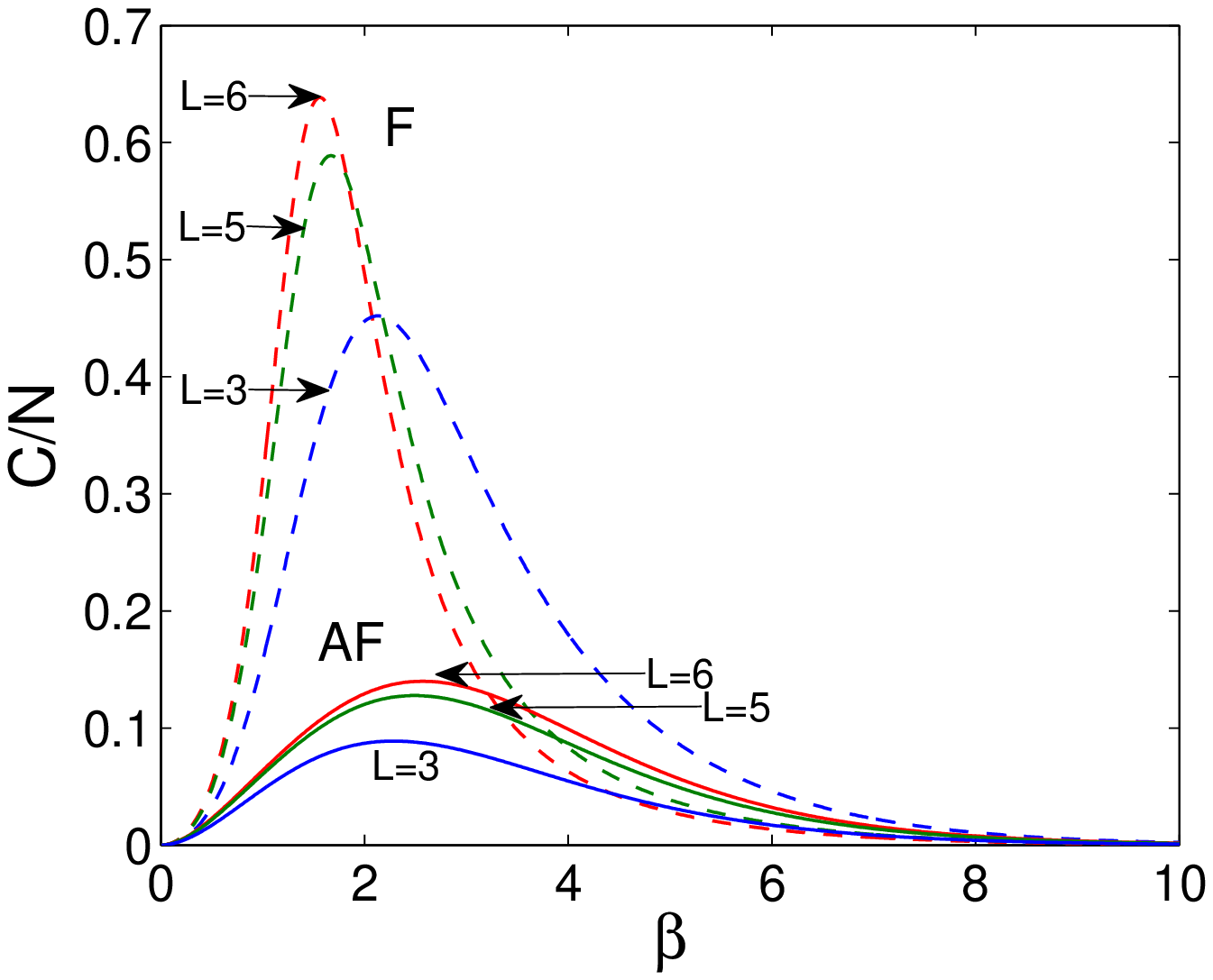}\label{fig:c-fin_T_T0}}
\caption{(a) Entropy and (b) specific heat per spin as functions of the inverse temperature $\beta$ for T ferromagnetic (F) and antiferromagnetic (AF) clusters of the side length $L$ and spin $s=1/2$. }\label{fig:ent-c_fin_T_T0}
\end{figure}

\begin{figure}[t]
\centering
    \subfigure[]{\includegraphics[scale=0.5,clip]{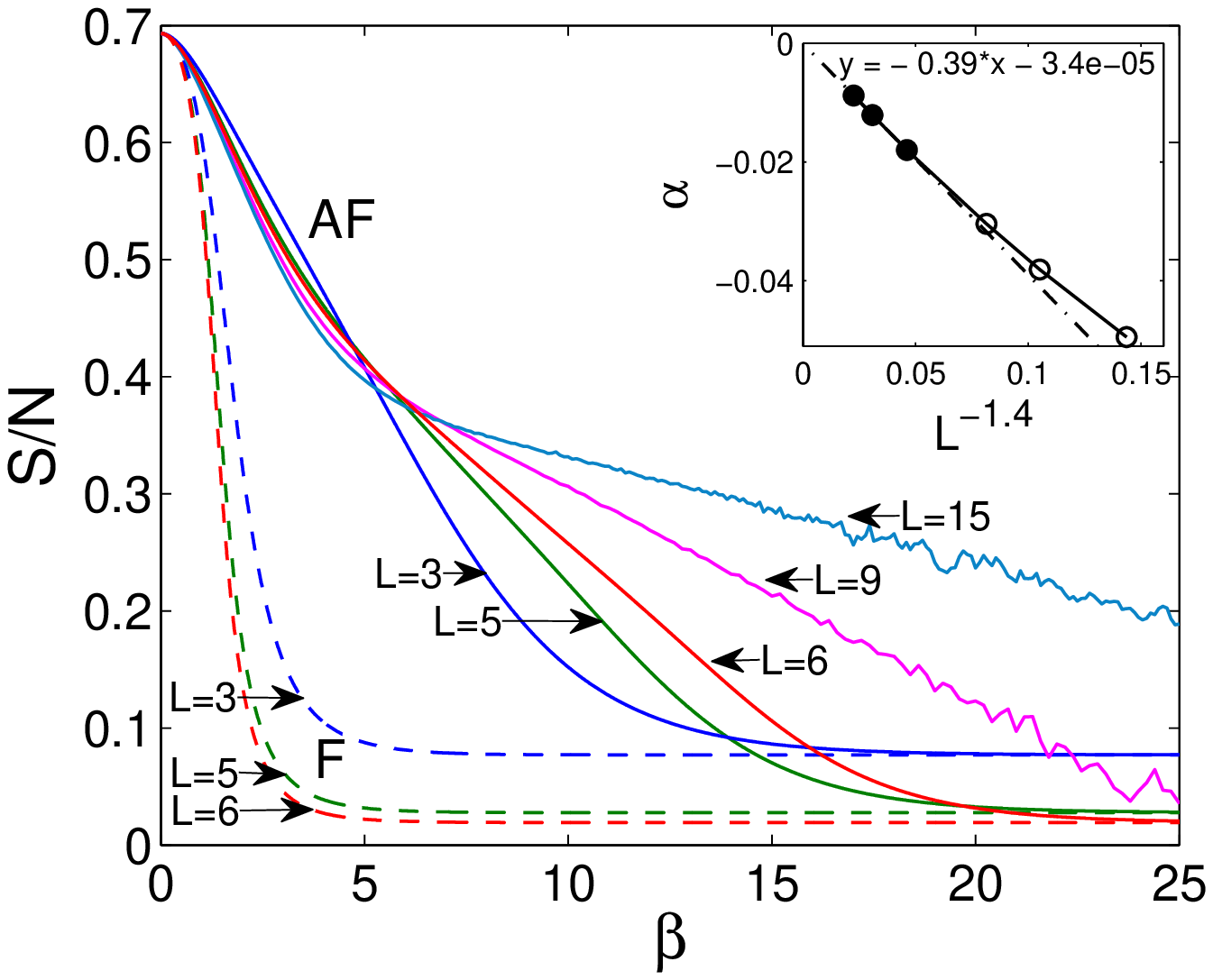}\label{fig:entr_rhomb}}
    \subfigure[]{\includegraphics[scale=0.5,clip]{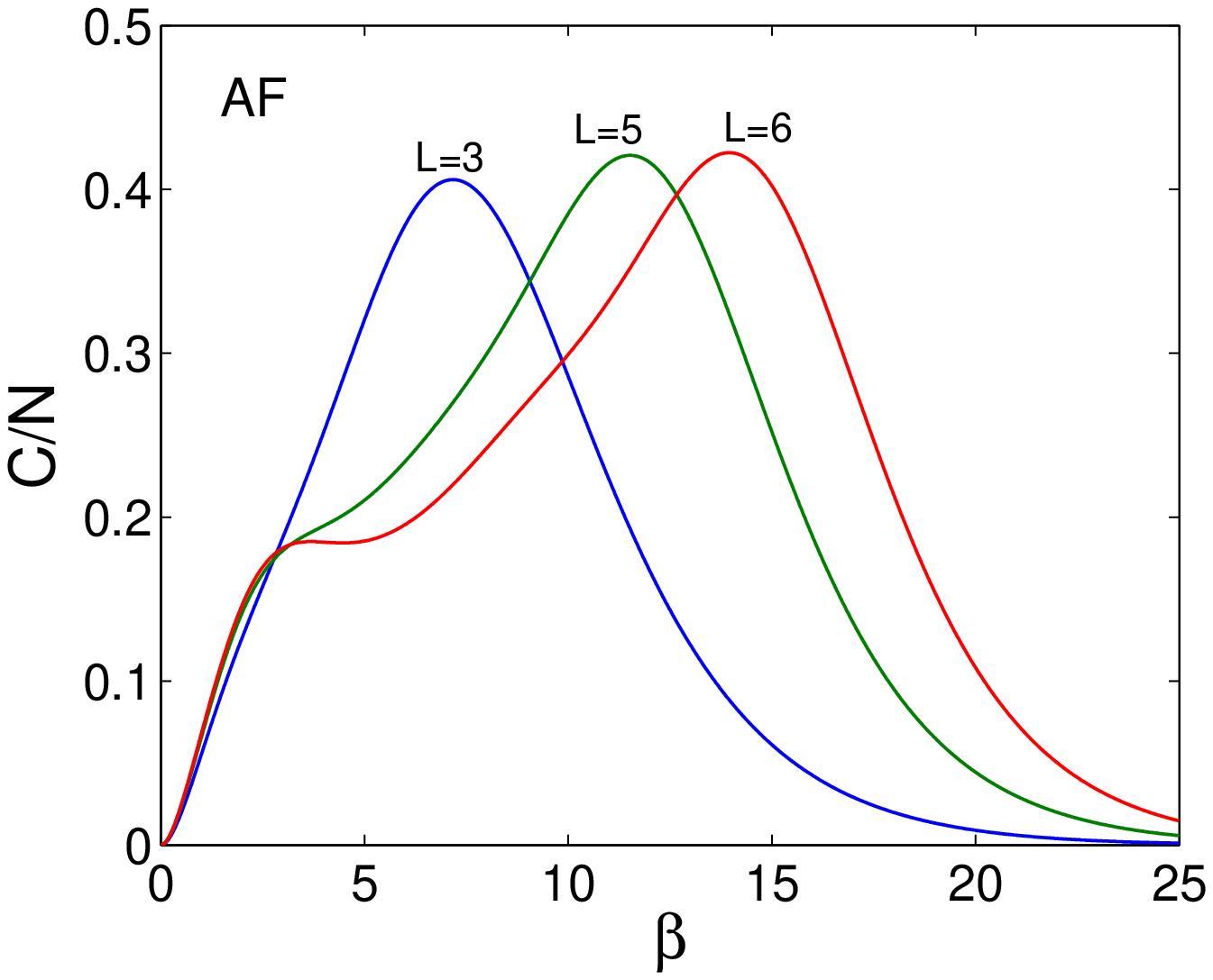}\label{fig:c_rhomb}}
\caption{(a) Entropy per spin as a function of the inverse temperature $\beta$ for R ferromagnetic and antiferromagnetic clusters with spin $s=1/2$. The cases of $L=9$ and $12$ were obtained from Monte Carlo simulation. The inset shows the slopes of the linear parts of the curves vs. $L^{-1.4}$ and the dashed line represents the best linear fit for $L \geq 9$. (b) Specific heat per spin corresponding to the antiferromagnetic clusters with $L=3,5$ and 6.}\label{fig:ent-c_rhomb}
\end{figure}

Exhaustive scanning of the entire state space allows an exact calculation of different thermodynamic functions also at finite temperatures. In Fig.~\ref{fig:ent-fin_T_T0} we show how the entropies of the T clusters approach their residual values for various cluster sizes. To focus on the behavior in the most interesting low-temperature region we plot the values as functions of the inverse temperature $\beta=1/T$ (solid curves) and to see the frustration effect we also include the results for the corresponding non-frustrated ferromagnetic (F) clusters (broken curves). The decay for the antiferromagnetic (AF) clusters is slower than for the ferromagnetic ones, due to larger residual values of the former, but otherwise they look similar. Namely, a sharp monotonic decrease occurs at moderate temperatures ($\beta \approx 2$) and the residual value is achieved already well above the ground state ($\beta \approx 10$). Nevertheless, some qualitative difference in the behavior of the anomalies related to the sharp entropy decrease between AF and F clusters can be seen looking at the corresponding specific heat curves, shown in Fig.~\ref{fig:c-fin_T_T0}. While with the increasing cluster size the peaks of the F systems move towards higher temperatures (the critical value of $\beta_c=\ln(3)/4$ in the thermodynamic limit~\cite{hout}), for the frustrated AF clusters the peak moves toward lower temperatures (from our MC simulations $\beta \approx 3.2$ in the thermodynamic limit).
\\
\hspace*{5mm} 
Similar behavior can also be observed in the remaining cluster shapes, except for the rhombus, the entropies of which are presented in Fig.~\ref{fig:entr_rhomb}. As expected from the ground-state discussion, at sufficiently low temperatures the curves of F and AF clusters of a given size merge and tend to zero as the cluster size increases. Nevertheless, the entropy values of AF R clusters do not approach the zero-temperature values in the same manner as those, for example, for AF T clusters. More specifically, after the initial sharp decrease the $\beta$-dependence changes to linear for a range of temperatures before the residual value is reached. The slopes of the linear parts of the curves become more gentle with the increasing cluster size and the dependence appears to be power law. In the inset of Fig.~\ref{fig:entr_rhomb} the slopes  $\alpha$ determined from the exact calculations for $L=3,5,6$ (open circles) and from MC simulations by applying the thermodynamic integration method~\cite{roma,zuko2} for $L=9,12,15$ (filled circles) are plotted against $L^{-1.4}$. Apparently the linear regime is reached only for sufficiently large cluster sizes. The dotted line represents the best linear fit based on the sizes $L \geq 9$. The thermodynamic limit extrapolation suggests that the zero entropy is reached only in the ground state, where AF and F curves merge. This would indicate much higher sensitivity of this non-degenerate ground state to thermal fluctuations, compared with both F and AF infinite-lattice systems for which the ground-state manifolds seem little affected for some rage of temperatures above zero~\cite{wann}. As a result, the specific heat curves for R clusters (see Fig.~\ref{fig:c_rhomb}) show, besides the higher-temperature anomaly, another low-temperature peak which moves to $\beta \to \infty$ for $L \to \infty$.\\
\begin{figure}[t]
\centering
    \subfigure[]{\includegraphics[scale=0.5,clip]{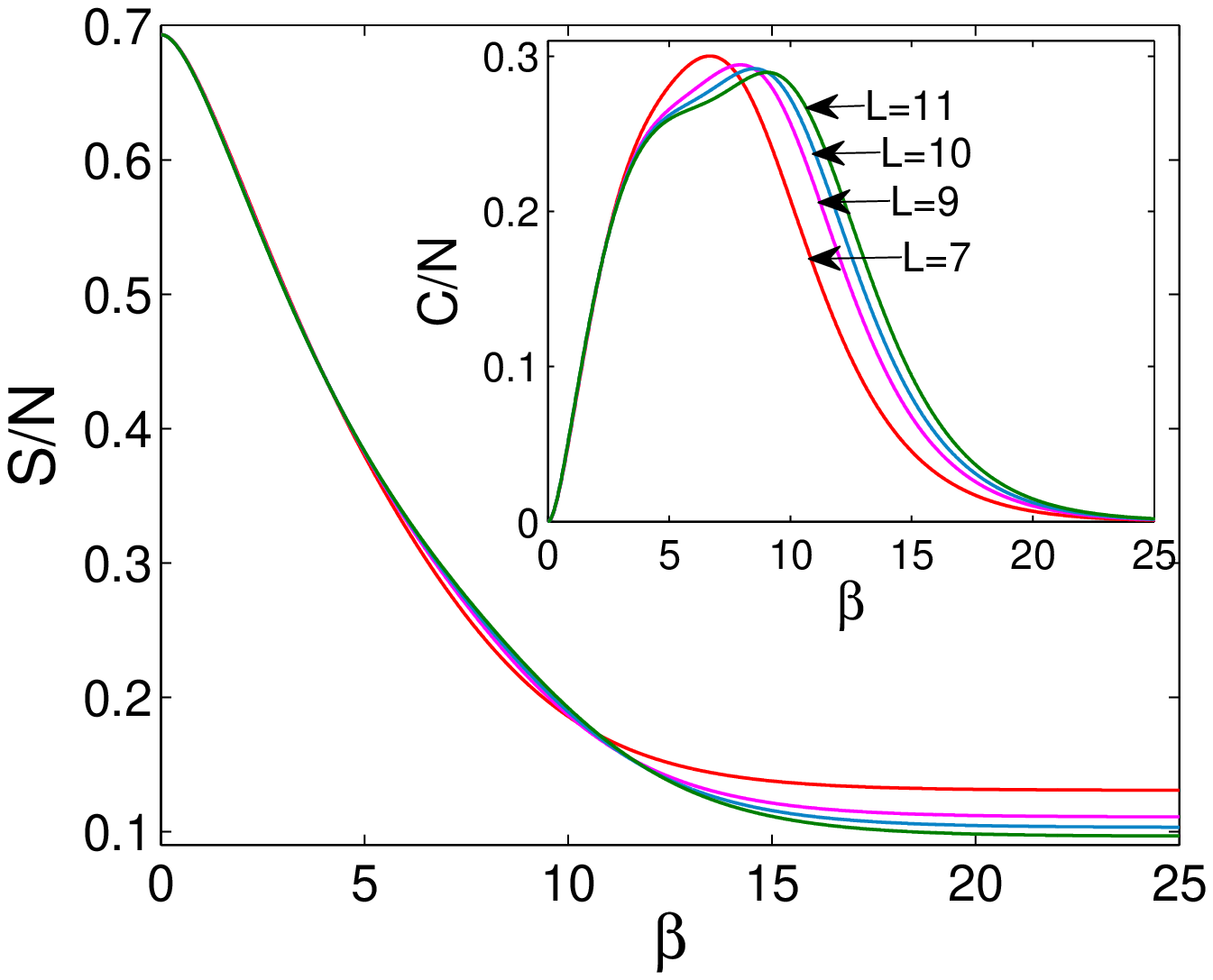}\label{fig:ent_c_chain1}}
    \subfigure[]{\includegraphics[scale=0.5,clip]{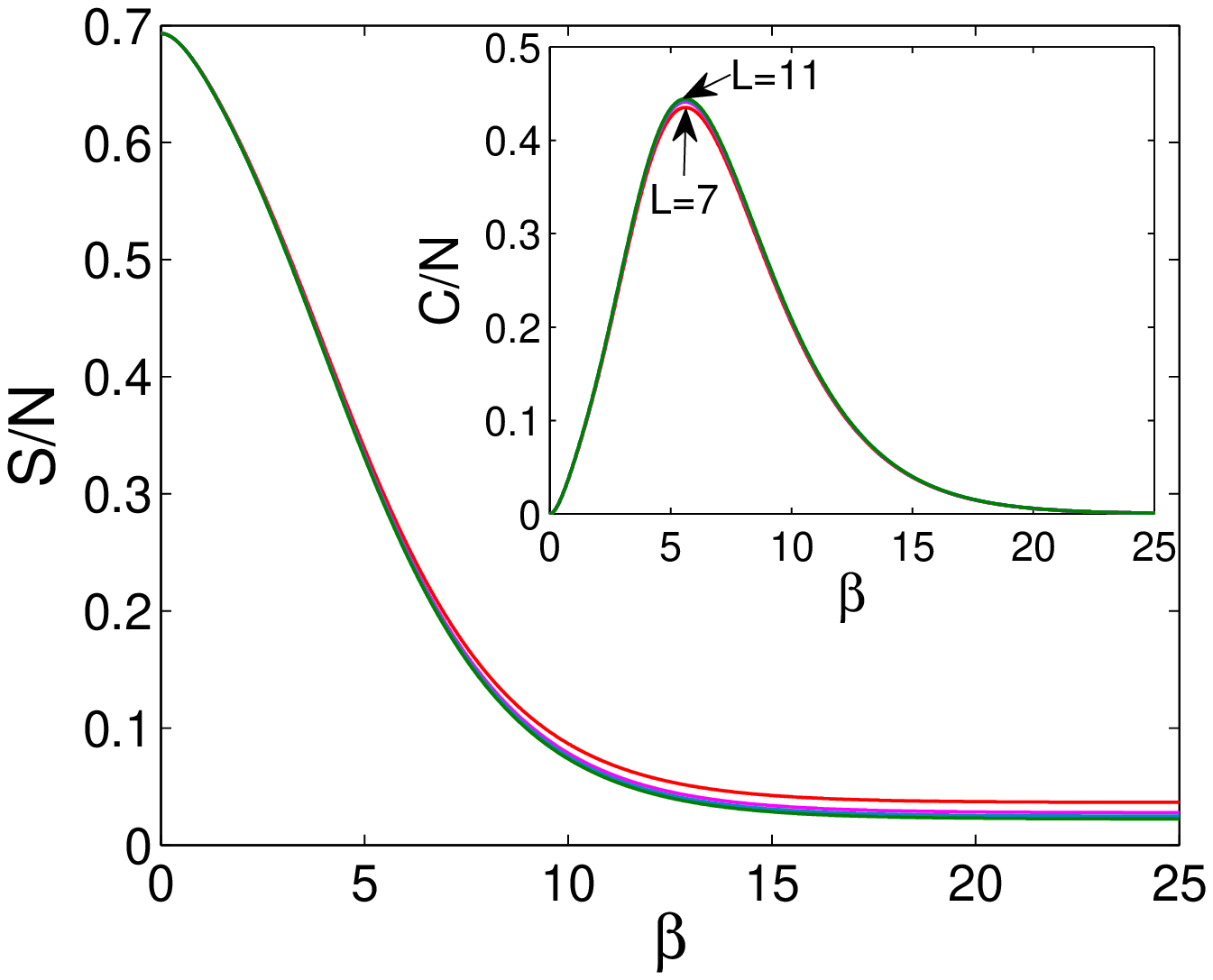}\label{fig:ent_c_chain2}}
\caption{Entropy and specific heat per spin (inset) as functions of the inverse temperature $\beta$ for AF (a) L1 and (b) L2 clusters with $L=7-11$ and spin $s=1/2$.}\label{fig:ent_c_chains}
\end{figure}
\hspace*{5mm} Similar anomalies can also be observed in the L1 clusters, albeit much less pronounced (Fig.~\ref{fig:ent_c_chain1}). On the other hand, only one specific heat peak can be seen in the L2 clusters and the curves almost collapse on each other for the considered sizes (Fig.~\ref{fig:ent_c_chain2}), even though they have no ground-state degeneracy just like the R clusters.

\section{Conclusions}
We studied effects of the shape, size and the spin values on the ground-state energy and entropy in geometrically frustrated finite clusters. We considered several shapes, derived from a rhombus on a triangular lattice by removal of some number of spins, and found substantial differences in the behavior of the studied quantities. With the increasing cluster size the reduced values of the ground state energy for all shapes and spin values converge to the same value of $-1$, however, the convergence rate differs among the shapes and for some shapes (such as T3) the cluster size dependence can be a non-monotonic function. On the other hand, the residual entropy for some cluster shapes, such as the triangle-based (T,T1,T2,T3) and the hexagon-based (H,H1), appears to converge to the finite values corresponding to the residual entropies of the infinite systems for a given spin value, while for the rhombus-based (R,R1) and ladder-like (L1,L2) shapes it vanishes as the cluster size increases. Like in the case of the energies, the dependences can be non-monotonic and the convergence rates strongly shape-dependent. The relative values normalized by the infinite-limit residual entropies generally decrease with the increasing spin value.\\
\hspace*{5mm} 
Finite-temperature calculations for the spin $s=1/2$ pointed to qualitative differences among different cluster shapes. Particularly interesting are the shapes the ground state degeneracy of which vanishes in the thermodynamic limit, i.e., the rhombus-based and ladder-like shapes. For example, even though the R clusters are known to have non-degenerate ground state, degeneracy is induced by thermal fluctuations at much lower temperatures (in the thermodynamic limit at any finite temperature), compared with some other shapes or non-frustrated clusters. Then for some temperature range the entropy increases linearly with the temperature before a sharp increase to the infinite-temperature value of $S/N=\ln(2)$. This behavior is reflected in two anomalies in the temperature dependence of the specific heat. \\
\hspace*{5mm} Finally, it is worth noticing that in the present frustrated spin clusters considerable entropy changes can be achieved by minimal spin manipulations. Remarkable examples include R $\leftrightarrow$ R1 transition by switching the state of a singe central atom between magnetic and non-magnetic or R $\leftrightarrow$ T transition by switching the states of the diagonal atoms in the rhombus between magnetic and non-magnetic, etc. Systems showing large isothermal entropy changes are important in the study of the magnetocaloric effect and search of efficient magnetic refrigerants. Therefore, it will be interesting to extent these investigations focusing on the magnetocaloric properties at finite temperatures.

\section*{Acknowledgments}
This work was supported by the grant of the Slovak Research and Development Agency under the contract No. APVV-0132-11.

\end{document}